\documentclass[tighten,times]{aastex631}
\usepackage{amsmath}
\usepackage{hyperref}

\newcommand{\SH}{Solberg-Høiland\ }

\newcommand{\BVF}{Brunt-V\"ais\"al\"a-frequency\ }
\newcommand{\BVFs}{Brunt-V\"ais\"al\"a-frequencies\ }

\received{April 27th, 2022}
\revised{April 1st, 2023}
\accepted{May 8th, 2023}
\submitjournal{ApJ}

\shorttitle{Thermal instabilities in accretion disks II: numerical experiments}
\shortauthors{Klahr et al.}

\begin{document}


\title{Thermal instabilities in accretion disks II:\\
Numerical Experiments for the Goldreich-Schubert-Fricke Instability and the Convective Overstability in disks around young stars}


\correspondingauthor{Hubert Klahr}\email{klahr@mpia.de}

\author[0000-0002-8227-5467]{Hubert Klahr}
\affil{Max Planck Institut f\"ur Astronomie, K\"onigstuhl 17, 69117, Heidelberg, Germany}
\author[0000-0002-0880-8296]{Hans Baehr}
\affil{Department of Physics and Astronomy, The University of Georgia, Athens, GA 30602, USA}
\affil{Center for Simulational Physics, The University of Georgia, Athens, GA 30602, USA}
\author[0000-0002-1697-6433]{Julio David Melon Fuksman}
\affil{Max Planck Institut f\"ur Astronomie, K\"onigstuhl 17, 69117, Heidelberg, Germany}




\begin{abstract}
The linear stability analysis of a stratified rotating fluid (see paper I) showed that disks with a baroclinic stratification under the influence of thermal relaxation will become unstable to thermal instabilities. One instability is the Goldreich-Schubert-Fricke instability (GSF), which is the local version of the Vertical Shear Instability (VSI) and the other is a thermal overstability, the Convective Overstability (COS). 

In the present paper we reproduce the analytic predicted growth rates for both instabilities in numerical experiments of small axisymmetric sections of vertically isothermal disks with a radial temperature gradient, especially for cooling times longer than the critical cooling time for VSI. In this cooling time regime our simulations reveal the simultaneous and independent growth of both modes: COS and GSF. We consistently observe that GSF modes exhibit a faster growth rate compared to COS modes. Near the midplane, GSF modes eventually stop growing, while COS modes continue to grow and ultimately dominate the flow pattern. 

Away from the midplane, we find GSF modes to saturate, when bands of constant angular momentum have formed. In these bands we observe the formation and growth of eddies driven by the baroclinic term, further enhancing the velocity perturbations. In geophysics this effect is known as horizontal convection or sea-breeze instability. Three-dimensional simulations will have to show whether similar effects will occur when axisymmetry is not enforced. Our local simulations help to reveal the numerical resolution requirements to observe thermal instabilities in global simulations of disks around young stars.
\end{abstract}


\noindent


\section{Introduction}
The planet forming dusty disks around young stars are subject to a range of magnetic and non-magnetic instabilities \citep{Lesur2022}.
The turbulence emerging from these instabilities has a profound impact on the planet formation process, via transporting and mixing dust, generating collisions among grains and ultimately influencing the migration of planets.
Among the pure hydro instabilities there are non-linear instabilities like the Zombie Vortex Instability (ZVI) \citep{Marcus2015}, the stratorotational instability \citep{Shalybkov2005} and the subcritical baroclinc instability (SBI) \citep{Petersen2007a}. The Rossby wave instability is a linear yet {a{radially global instability} \citep{Lovelace1999} and likewise the vertical shear instability (VSI) \citep{Urpin1998,Nelson+2016} is a linear {vertically global instability}.  

The latter instability is actually an extension of the Goldreich-Schubert-Fricke instability (GSF) as studied in rotating stars \citep{Goldreich1967,Fricke1968} for the geometrically thin accretion disk. 
GSF operates the best for short cooling times, but operates actually for any cooling rate \citep{Tassoul2000}, albeit at slower growth rates as discussed in \citet{Urpin2003}. 
The original VSI work \citep{Urpin1998,Urpin2003} considers local modes, whereas all recent analytic work on VSI considers vertical global modes \citep{Nelson+2016,Lin2015,Barker2015,Latter2017,Cui2022,Latter2022}. So to separate the local treatment in the present paper from the global treatment in these papers, {we refer} to the GSF, by which {we mean} discussing growth rates from the local dispersion relation including thermal relaxation \citep{Klahr2023}. By this definition GSF is a local instability, whereas VSI is an overstability, even so the stability criterion is the same.
In \citet{Tassoul2000} the GSF is classified as a thermal instability, which together with the "Vibrational Instability of Rotating Stars" by \citet{Shibahashi1980} (aka a thermal overstability) forms a set of "thermal instabilities".

The Shibahashi process was rediscovered for disks around young stars by \citet{Klahr2014} and dubbed convective overstability. Initial studies neglected the vertical stratification \citep{Lyra2014}, yet lately it was shown that COS will also exist in a vertically stratified disk \citep{Klahr2023} (Paper I). This paper predicts growth rates for both GSF and COS modes as a function of disk stratification and cooling rate. The formalism to determine these growth rates was already derived in \citet{Urpin2003}, yet the existence of convective modes was not further investigated. In Paper I it was finally shown that a baroclinic atmosphere with its non-parallel contours of pressure and entropy must also {possess} directions in which the stratification is super-adiabatic. While this super-adiabatic stratification does not directly lead to convection as discussed in the \SH criterion \citep{Ruediger2002}, because of the stabilizing effect of the epicyclic term in the absence of thermal relaxation, it is sufficient to amplify epicyclic oscillations for a thermal relaxation time on the order of the epicyclic oscillation time, i.e.\ the Keplerian period.

Also the GSF \citep{Urpin2003} and the VSI \citep{Lin2015} {possess} growth rates in this cooling time regime beyond the critical cooling time $\tau_c$ for VSI. Yet, numerical studies were not able to reproduce them for global simulations of disks \citep{Manger2021}. The VSI seemed to be suppressed in the case of cooling times longer than the critical cooling time.

\begin{deluxetable}{lll}[tb!]
\tabletypesize{\footnotesize}
\tablecaption{Used symbols and definitions:\label{tab:usedSym}}
\tablehead{\colhead{Symbol} & \colhead{Definition }& \colhead{Description}}
\startdata
$R$,$z$,$\phi$         &                        & cylindrical coordinates   \\
$\rho, P$      &            & density, pressure\\
$c_v$      &                          & specific heat \\
$E, T$      &  $E = c_v \rho T$          & internal energy, temperature\\
$\gamma$      &                          & adiabatic index \\
$K$      &      $P \rho^{-\gamma}$      & specific entropy\\
$p$         &     $ = \frac{ {\rm d\log}\,\rho(R,0)}{{\rm d\log} R}  $  & global density gradient \\
$q$         &    $ = \frac{{\rm d \log} \, T(R,0)}{{\rm d \log} R}     $ & global temperature gradient\\
$a_R,a_z$     &    $ \nabla {\rm log} \rho   $ & local density stratification\\
$b_R,b_z$     &    $ \nabla {\rm log} P   $ & local pressure stratification\\
$s_R,s_z$     &    $ \nabla {\rm log} K   $ & local entropy stratification\\
$\Omega$      &            & Keplerian frequency\\
$c$      &      $P / \rho$        & isothermal speed of sound\\
$H$      &          $c / \Omega$                     & pressure scale height\\
$h$      &          $H/R$                     & aspect ratio\\
$\tau$      &                                 & thermal relaxation time\\
$\tau_c$      &                               & critical $\tau$ for VSI\\
$\tau^*$      &      $\tau \gamma \Omega$                           & dimensionless cooling time\\
$k_R,k_z$  &                          & radial, vertical wave number\\
$\mathbf{k}$  &                          & wave number vector\\
$\mathbf{a}$  &    $\mathbf{a}\cdot \mathbf{k} = 0$                     & direction vector for velocity perturbation\\
$\kappa_R^2$      &       $=\frac{1}{R^3}\partial_R \Omega^2 R^4$                    & radial angular momentum gradient: epicyclic frequency \\
$\kappa_z^2$      &         $=\frac{1}{R^3}\partial_z \Omega^2 R^4$                  & vertical angular momentum gradient\\
$\kappa_\mathbf{k}^2$      &    $=\frac{k^2_z}{k^2}\left(\kappa_R^2 - \frac{k_R}{k_z}\kappa_z^2\right)$                       & oscillation frequency (OF) for $\mathbf{k}$\\
$N^2$      &    $= - \frac{1}{\rho \gamma c_v}\nabla P \nabla K$                & buoyancy frequency (BF)\\
$N_R^2,N_z^2$      &       & radial, vertical BF\\
     &                    & aka \BVF\\
$N_\mathbf{k}^2$      &   $=\frac{N_R^2 \left(1 -\frac{b_z k_R}{b_R k_z}\right) k_z^2 + N_z^2 \left(1 - \frac{b_R k_z}{b_z k_R}\right) k_R^2 }{k^2}$                 & buoyancy frequency for $\mathbf{k}$\\
$N_-^2$      &    $=\min(N_\mathbf{k}^2)_\mathbf{k}$                & lowest local BF\\
$N_+^2$      &    $=\max(N_\mathbf{k}^2)_\mathbf{k}$                & largest local BF\\
$\kappa_-^2$      &    $=\min(\kappa_\mathbf{k}^2)_\mathbf{k}$                & lowest local OF\\
$\kappa_+^2$      &    $=\max(\kappa_\mathbf{k}^2)_\mathbf{k}$                & largest local OF\\
$\Gamma_\mathrm{VSI}$      &    $=\frac{H}{R}|q| \Omega$                & typical VSI growth rate\\
$\Gamma_\mathrm{GSF}(\tau < \tau_c)$      &    $=\frac{|z|}{R} \frac{|q|}{2}\Omega$                & GSF growth rate for $\tau \rightarrow 0$\\
$\Gamma_\mathrm{GSF}$      &    $=\frac{q^2}{4}\frac{\gamma }{\gamma-1}\,\,\, \frac{1}{\tau^*_\mathrm{GSF}  + \tau^*} \, \Omega$                & approx. GSF growth rate\\
$\Gamma_\mathrm{COS}$      &    $=\frac{q^2}{8}\frac{\gamma}{\gamma-1}\,\, \frac{\tau^*}{1 + \tau^{*2}} \,\Omega$                & approx COS growth rate\\
\enddata
\end{deluxetable}
The growth rates of the fundamental large-scale VSI modes \citep{Nelson2013} scale proportional to the disk aspect ratio $h = H/R$, radial temperature stratification $q$ and Keplerian frequency $\Omega$
\begin{equation}
   \Gamma \approx h |q| \Omega,
\end{equation}
which can be reproduced in numerical simulations \citep{Nelson2013,Stoll2014,Richard2016,Stoll2017,Manger2021} as long as the cooling time is shorter than the critical value $\tau_c$ derived by \citet{Lin2015} (see also the local variant in \citet{Urpin2003}). 
In the present paper we show that, with sufficient resolution of limited radial and vertical extent, we can reproduce the predicted growth-rates of Paper I. Translating our local resolution of 256 cells per pressure scale height $H$ to a global simulation covering $\pm 3.5 H$ would need 1792 cells in the vertical direction to reproduce the growth of GSF and COS modes for long cooling times. 
 
{Both GSF and COS modes can only be avoided in a barotropic atmosphere, i.e.\ if pressure is a function of density only, which means either globally constant entropy or globally constant temperature in the disk.} Any temperature structure that is not barotropic (aka baroclinic), will lead to instability of both GSF and COS modes. For short cooling times $\tau < \tau_c$ 
GSF will dominate, but for longer cooling times both GSF and COS have very similar growth rates. {Thus, in disks around young stars with thermal relaxation, both instabilities will always co-exist}.
For the purpose of a numerical experiment we will use a slightly non-physical gravity law to either suppress COS or GSF, but for conservative gravity, this is not possible.

Even in regions which are radially stably stratified (with respect to convection) one can observe the development of slanted COS modes. It is a major result of Paper I 
that COS does not strictly depend on radial unstable modes, but that slanted modes also play an important role. 

An effect of the limited vertical extent of our simulation domain can be observed in the saturation behavior of the linear growth phase. Whereas in the global simulations eventually Kelvin-Helmholtz eddies are created between the vertical VSI modes as shown by (Melon Fuksman et al. A\&A, submitted) we see a rather smooth formation of radial confined regions of constant angular momentum. {In these bands} there is no vertical shear and thus GSF growth does not exist anymore. Also the COS does not persist anymore as the radial epicyclic frequency vanishes. Convection as described in the \SH criterion does also not occur as we are able to measure. It is the baroclinic term itself that is now driving clockwise rotating eddies within the bands of constant angular momentum, similar to horizontal convection or the "Sea-breeze" mechanism in geophysics \citep{Holton2012-kg}. The mechanism is also related to the SBI \citep{Petersen2007a}, yet in contrast to that it operates also for very short cooling times. The SBI, on the other hand, operates in the barotropic background of the disk, with the baroclinicity only generated by the rotation of the in-plane vortex itself. This hysteresis effect only occurs when the cooling rate is of the order of the rotation frequency of the vortex \citep{Lesur2010,Raettig2012}. 

\subsection{Outline}
We test our predicted growth rates in non-linear hydrodynamic simulations for sufficiently small sections of a disk atmosphere (to narrow down the possible range of growth rates) in Section 2.1. In Section 2.2, we perform test simulations in which we suppress either COS or GSF by modifying stellar gravity to explore the unstable modes in their linear evolution and non-linear saturation independently. Finally, in Section 2.3, we give an example of the development of diagonal COS modes in radially stably stratified disks with a steep temperature gradient, as normal temperature gradients lead to insufficient growth rates to be handled with our dissipative numerical scheme.
In Section 3 we discuss eddies that we find in 
the non linear state of our simulations, amplified by a separate process from COS and VSI. We identify the driving process as horizontal convection similar to a "Sea-Breeze" in geophysics.
Section 4 summarizes our findings and gives an outlook to future work. Some details on the four movies we present along with this paper can be found in the appendix.
\begin{figure*}
   \centering
    \includegraphics[width=\linewidth]{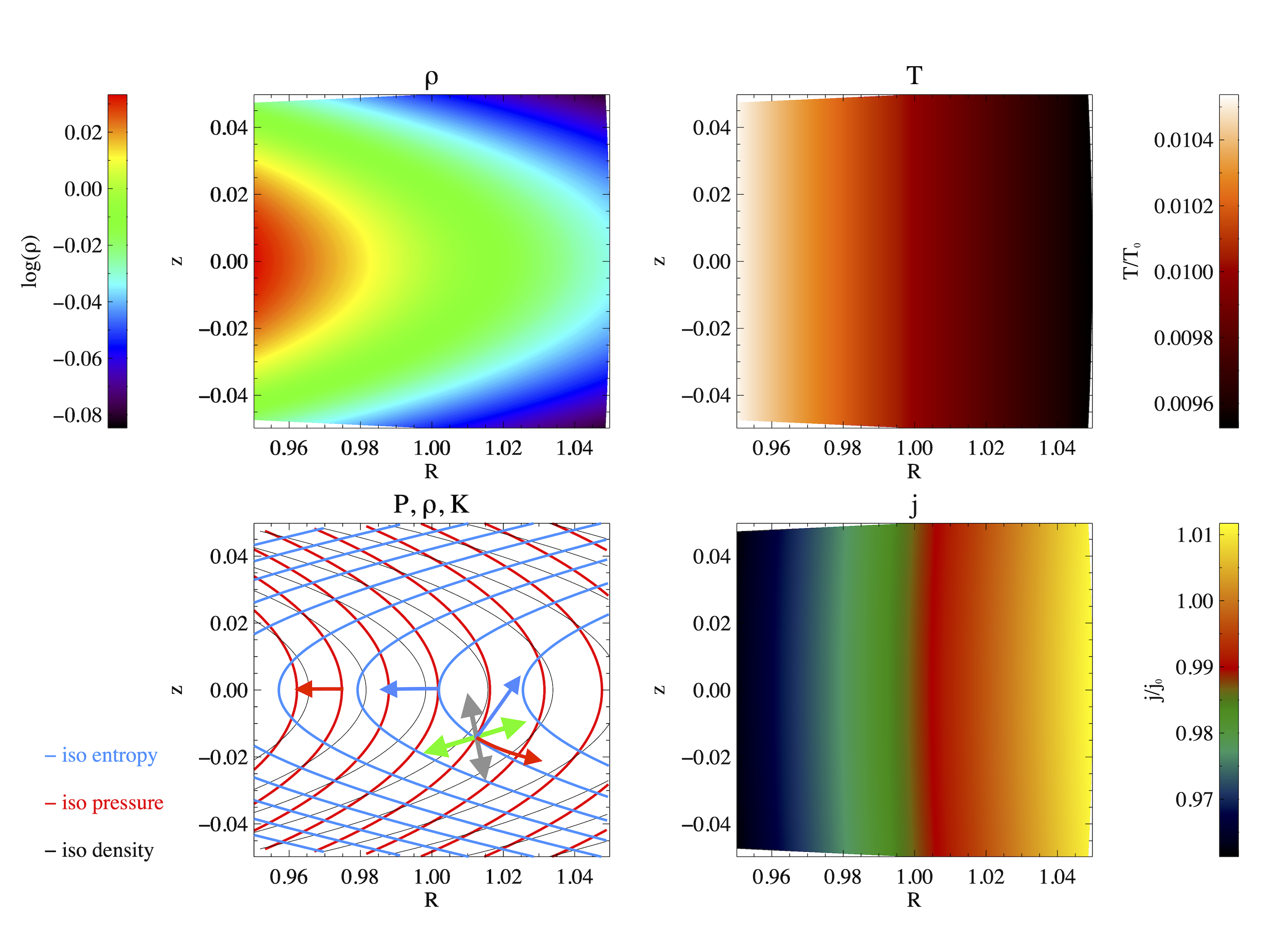}
    \caption{Background structure of the simulation area around the midplane ($z_0 =0$) for a disk with $q=-1$ and $p=-1.5$. Density $\rho$, temperature $T$, iso-contours and specific angular momentum $j = \Omega R^2$.
    We emphasize the iso-contours for pressure (red) and entropy (blue). An the disk midplane, both gradients in entropy and pressure point radially inward (red and blue arrow). Away from the midplane, the iso-contours and thus the gradients {bend} with respect to each other and the direction of largest unstable buoyancy (green arrows) is no longer strictly radial but points towards and away from the midplane. This is the direction in which the COS would operate. The opposite direction is stably stratified (grey arrows). The non-alignment of density and pressure (baroclinic structure) is the reason the specific angular momentum decreases with height.
    } 
        \label{Fig:mod2D256Q1_V4DW0}
\end{figure*}
\begin{figure*}
   \centering
    \includegraphics[width=\linewidth]{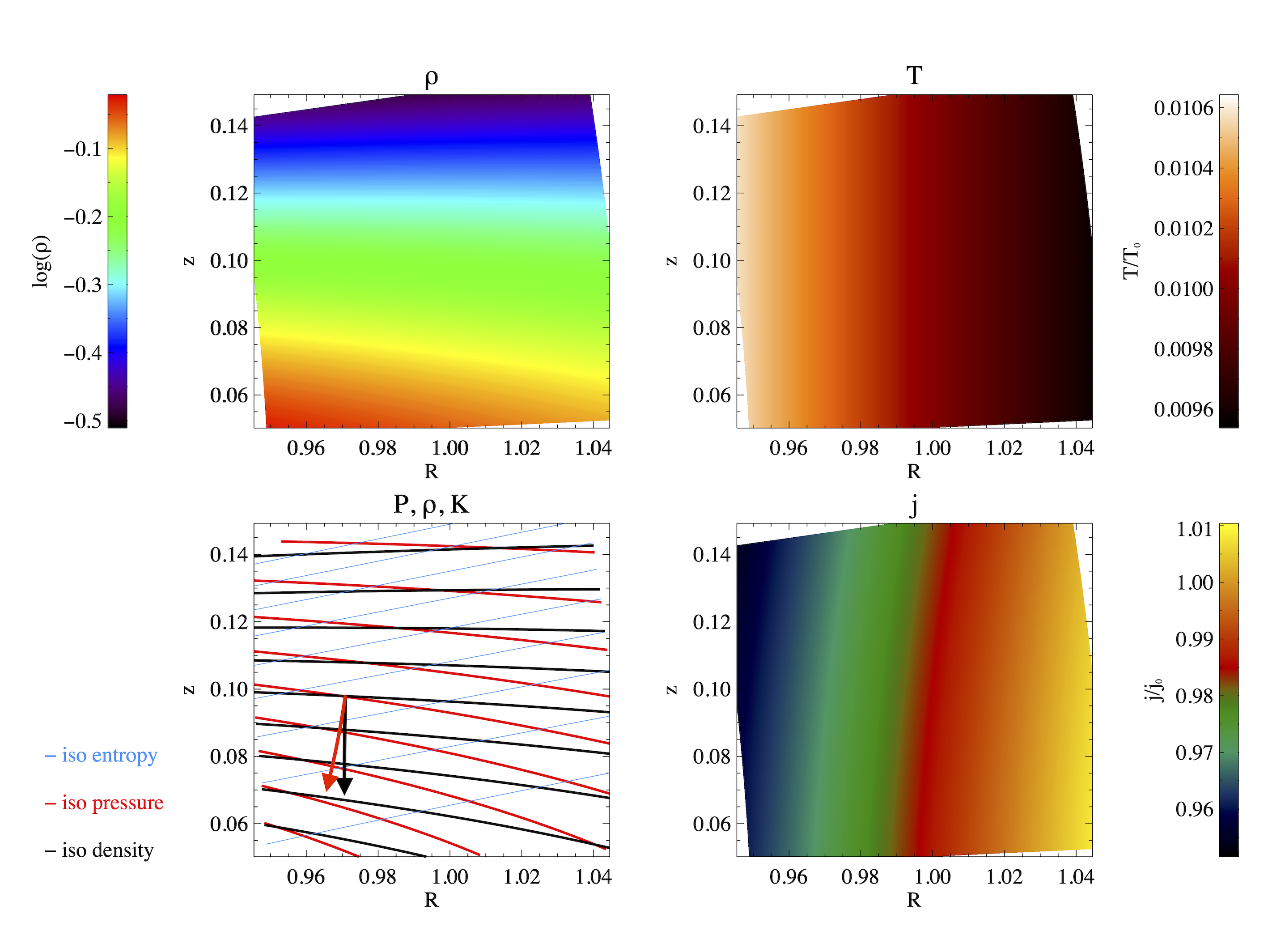}
    \caption{Background structure of the simulation area around one pressure scale height above the midplane ($z_0 =H$) for a disk with $q=-1$ and $p=-1.5$. Density $\rho$, temperature $T$, iso-contours and specific angular momentum $j = \Omega R^2$.
    We emphasize the iso-contours for density (black), pressure (red) and entropy (blue). At the midplane, both gradients in density and pressure point radially inward. Away from the midplane, the iso-contours and thus the gradients in density (black arrow) and pressure (red arrow) bend with respect to each other (baroclinic structure), which is the reason the specific angular momentum $j$ decreases with height, or respectively the reason its iso-contours bend outward, which is the cause of the GSF and VSI.} 
        \label{Fig:mod2D1H256Q1_V4DW0}
\end{figure*}
\begin{figure*}
   \centering
    \includegraphics[width=\linewidth]{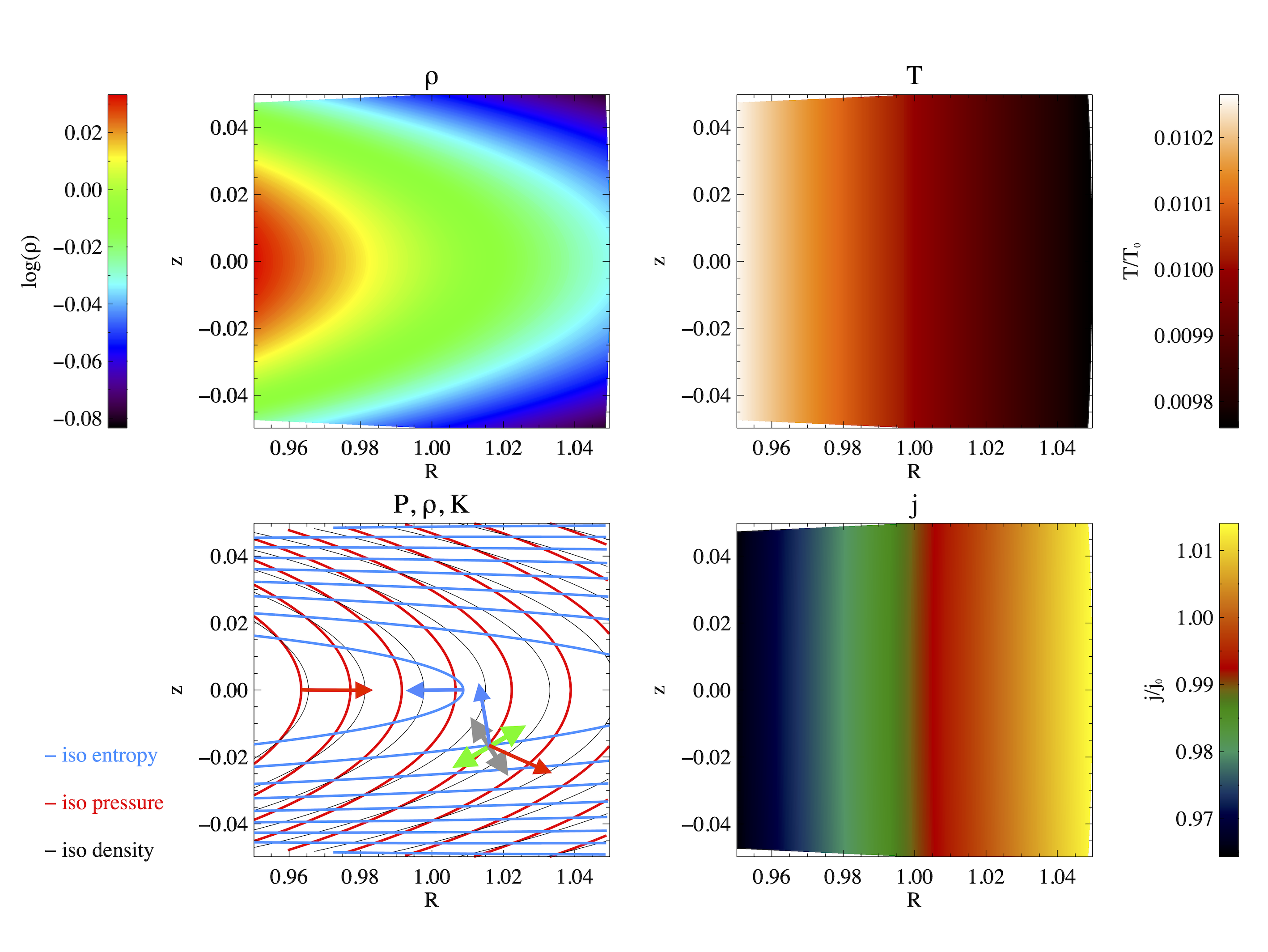}
    \caption{Background structure of the simulation area around the midplane ($z_0 =0$) for a disk with $q=-0.5$ and $p=-1.5$. Density $\rho$, temperature $T$, iso-contours and specific angular momentum $j = \Omega R^2$.
    We emphasize the iso-contours for pressure (red) and entropy (blue). Due to the shallower temperature gradient, the entropy gradient in the midplane (blue arrow) points {inward}, where as the gradient in pressure points radially {outward} (red arrow), thus the midplane is radially stable with respect to convective modes. Outside the midplane, the iso-contours and thus the gradients twist with respect to each other and they do not point strictly in opposite directions. Thus, there is now a direction of unstable buoyancy (green arrow) very similar to the case with $q=-1$. This is the direction in which the COS would operate. The opposite direction is stably stratified (grey arrows). The bending of the iso-contours in specific angular momentum as a cause for VSI and GSF modes is also present, only weaker due to the shallower temperature profile.} 
        \label{Fig:mod2D256Q05_V4DW0}
\end{figure*}

\section{Numerical Experiments}
In paper I, we have seen that even for a given local stratification and thermal relaxation time, more than one instability can grow. In order to disentangle the various mechanisms, vertical shear vs.\ super-adiabatic stratification, we perform a limited set of numerical experiments. 

Specifically we want to restrict ourselves to three key questions in the present paper:
\begin{enumerate}
\item{Can we reproduce the predicted growth rates in numerical simulations of stratified accretion disks using the full non-linear set of hydrodynamic equations?}
\item{Can we distinguish between COS and GSF modes in these simulations?}
\item{Can we reproduce the inclined COS modes in simulations with radially stable stratification?}
\end{enumerate}
For all these tests we will apply axisymmetric setups and also restrict the simulation area to just a fraction of the disk atmosphere. This will 1) allow for a controlled experiment as the growth rates are height ($z$) dependent and 2) allow us to ensure sufficient numerical resolution and a reasonable run time. 
This means we will not be able to address the full development of a saturated three dimensional turbulent state and the angular momentum transport associated with it. Initial tests indicate a large numerical expense for such a study, especially if one wishes to adopt realistic values for the radial temperature gradient. We shall postpone these simulations to a future paper.

The first question we will address by a set-up similar to \citet{Manger2021}, yet apply a much smaller computational domain.

\subsection{Initial and Equilibrium Disk Structure}

Using the same grid size and resolution as in our global simulations  \citep{Manger2020,Manger2021,Pfeil2021} does not provide the sufficient resolution to study the linear development of some of the slowly growing instabilities we consider in this paper. Furthermore, a global disk has a wide range of growth rates, depending on distance to the star as well as distance to the midplane. We will therefore pick two small sections of the {disk} roughly covering a vertical range either sitting in the midplane ($z_0 = 0$): $z = -0.5 H$ to $z = 0.5 H$ or in the atmosphere ($z_0 = H$): $z = 0.5 H$ to $z = 1.5 H$. The radial range will then span from $R = 1 - 0.5 H$ to $R = 1 + 0.5 H$ and an axisymmetric region of height and width $H$. This shall allow for the study the effect of stratification on the unstable modes without mixing too many different conditions for instability and growth rates in one simulation and thus we can compare growth rates with analytical predictions.

Using small computational domains on the order of a pressure scale height makes one wonder why not use shearing sheet coordinates as was successfully done for the the MRI (magneto rotational instability \citep{Balbus1991}). For the SBI (subcritical baroclinic instability) in \citet{Lyra2011} it was already necessary to introduce the global gradient of entropy in a linearized way. The same setup for the COS in \citet{Lyra2014} did not consider vertical stratification. The effect of vertical density stratification, in combination with radial temperature stratification, to lead to vertical shear would then also have to be incorporated by linearizing certain terms, which may even be prone to numerical artifacts once radial periodicity is applied \citep{McNally2015}.
Thus, in the spirit of \citet{Klahr2014}, which used a cylindrical yet vertically unstratified set up we now go straight for a "global" setup in terms of applied equations, but use a small simulation domain. 

We use the PLUTO code (version 4.3) \citep{Mignone2007} in a spherical (or cylindrical) axisymmetric setup using special reflective boundary conditions that invert the normal component of velocities at the boundaries (both radial and polar, respectively vertical). In a departure from normal reflective boundaries, we impose the initial values for all other quantities (pressure, density, rotation profile) for the other ghost-cell values. The radial velocity at the vertical boundaries is defined as free slip, i.e., there is zero gradient towards the ghost cells. We treat the vertical velocities at the radial boundaries in the same fashion. 
These boundaries ensure that we lose no mass and that even a strongly perturbed disk can decay towards the initial and equilibrium state.

For the other details of our simulations we refer to \citet{Manger2018}
In spherical coordinates, the small disk region is defined as $r_\mathrm{min} = 0.95 $ to $r_\mathrm{max} =  1.05$ in radius, and from either A: $\theta_\mathrm{min} = \frac{\pi}{2} - 0.5 h$ to $\theta_\mathrm{max} = \frac{\pi}{2} + 0.5 h$ 
or B: $\theta_\mathrm{min} = \frac{\pi}{2} + 0.5 h$ to $\theta_\mathrm{max} = \frac{\pi}{2} + 1.5 h$.
The standard resolution is $256^2$ cells, i.e.\ 256 cells per pressure scale height. Lower resolution than this weakens the growth rates significantly. 
For a few cases ($q=-0.5$), we doubled the resolution to 512 cells per pressure scale height for better convergence towards the predicted growth rates.

The boundary conditions could in principle affect our simulations, but at least in the linear and axisymmetric regime, we find no severe impact of the quasi-reflective boundary conditions on the evolution of perturbations. {We tested that in simulations without thermal relaxation as well as with thermal relaxation, but no temperature gradient, and found that perturbations always decayed.}

During the non-linear stage of the instability we start to see some reflection of waves at the boundaries, which implies that for larger scales and especially full three dimensional simulations one will have to introduce damping layers as in \cite{Manger2018}.

We assume that our disks are vertically isothermal, such that the slope of midplane density $p$ and temperature $q$, as well as the aspect ratio $h_0 = H/R$ at radius $R_0$ defines our initial and background state of temperature $T(R,z)$ and density $\rho(R,z)$ in our simulations {as}
\begin{equation}
\rho(R,0)=\rho_0\left(\frac{R}{R_0}\right)^{p}, \,\,\,\,\,\,\,\,\, T(R,0) = T_0\left(\frac{R}{R_0}\right)^{q}.
\end{equation}
The vertical density structure is then
\begin{equation}
\rho(R,z) = \rho(R,0) e^{\frac{R^2}{H^2}\left(\frac{R}{\sqrt{R^2 + z^2}} - 1\right)},
\end{equation}
and the equilibrium rotation profile is 
\begin{equation}
    \Omega^2 = \frac{M G}{R^3}\sqrt{1 + q \left(1 -  \frac{R}{\sqrt{R^2  + z^2}}\right) + (p + q) \frac{H^2}{R^2}}.
\end{equation}
Our simulations are dimension free thus gravity constant $G$ and stellar mass $M$ are both equal to $1$. Then the Keplerian speed at radius $R = 1$ is also $1$, as is the Keplerian frequency $\Omega_\mathrm{Kepler} = 1$, which defines the orbital period as $t_\mathrm{Orbit} = 2 \pi$.

In Figure \ref{Fig:mod2D256Q1_V4DW0} and Figure \ref{Fig:mod2D1H256Q1_V4DW0} we show the initial density and temperature structure for model A with a radial temperature gradient of $q=-1$ and density gradient $p=-1.5$. Note the non-alignment of density, pressure and entropy contours, which constitutes the baroclinic state of the disk. This baroclinicity is likewise the cause for vertical shear and for convectively unstable directions. Note that neither {the GSF nor VSI} have strictly vertical modes, nor does COS have strictly radial modes. Nevertheless, GSF and COS modes are typically orthogonal to each other.
In Figure \ref{Fig:mod2D256Q1_V4DW0}, we emphasize that even close to the midplane, convective modes are not strictly radial, and the vertical shear is weaker than in the atmosphere as can be seen in Figure \ref{Fig:mod2D1H256Q1_V4DW0}. 

Our models with a shallower temperature gradient $q=-0.5$ are convectively stable in the midplane, as the entropy increases with distance to the star, but the pressure drops (See Figure \ref{Fig:mod2D256Q05_V4DW0}). Outside the midplane, the pressure gradient turns towards the midplane, whereas the entropy gradient turns away from the midplane. As a result, outside the midplane, there are directions which are convectively unstable in the spirit of the COS.
We omit a figure for the same parameters for the upper atmosphere as it looks qualitatively like the model with the steeper temperature gradient.

\begin{figure*}
    \centering
       \gridline{\fig{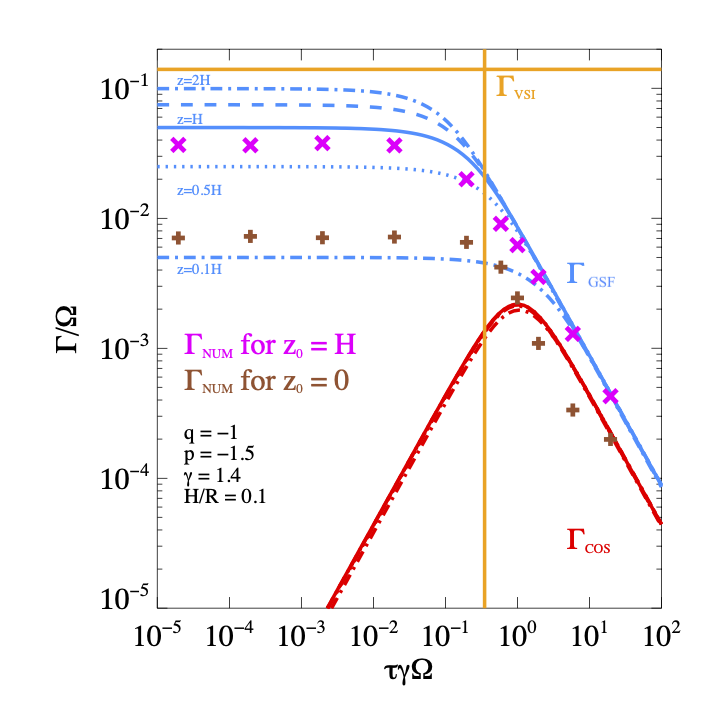}{0.5\textwidth}{(a): q = -1}\fig{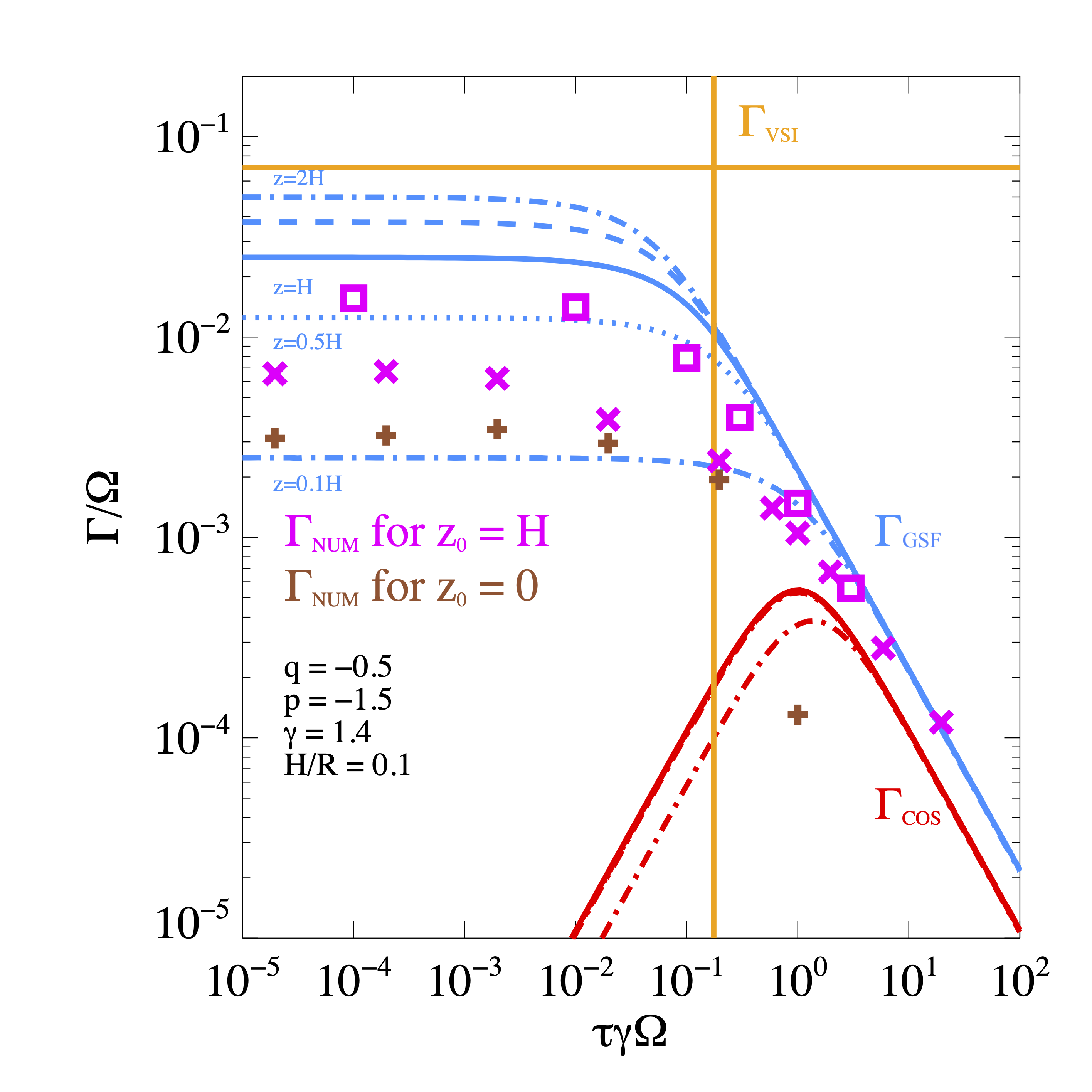}{0.5\textwidth}{(b): q = -0.5}}
    \caption{Numerically determined growth rates compared to analytic growth rates $\Gamma$ for COS and GSF for $p = -1.5$ as function of cooling time $\tau$ for various heights above the midplane. (a): $q = -1$ and (b): $q = -0.5$. In ascending order $z =  0.1H, 0.5H, H, 1.5 H, 2 H$ for a disk with $H/R = 0.1$ and the adiabatic index of $\gamma = 1.4$. The {solid} blue line corresponds to $z=H$. The strength of COS is mostly independent of height, which makes the red lines almost indistinguishable. The brown symbol $+$ are measured growth rates in local axisymmetric hydrodynamic simulations (at 256/$H$) for a box centered around $z_0=h$ and magenta $X$ symbols the same for a box at $z_0=0$. For $q = -0.5$ we added some runs at double resolution (512/$H$), indicated by magenta squares, which produces a better agreement.} 
        \label{Fig:TAUGRATE4Q1}
\end{figure*}

\subsection{Growth Rates: Local spherical simulations}
The full growth rates for COS and GSF have been determined by solving the dispersion relation from Paper I, which is equivalent to the one in the paper by \citet{Goldreich1967} as well as in the book by \citet{Tassoul2000}:
\begin{eqnarray}
\omega ^3  + \omega ^2  \frac{i}{\gamma  \tau} - \omega \left[N_\mathbf{k}^2 + \kappa_\mathbf{k}^2\right] - \frac{i}{\gamma  \tau } \kappa_\mathbf{k}^2 = 0.
\label{eq:EasyDisp}
\end{eqnarray}
by numerical methods, which involves identifying the optimum wavenumber vector $\mathbf{k} =(k_R,k_z)$ for either COS or GSF.
Additionally analytic approximation have been given is Paper I, as was already suggested by \citep{Urpin2003}.
Stability and growth rates are determined by the oscillation frequency\footnote{This is equal to $Q^2$ in \citet{Urpin2003}. See our discussion in Paper 1.} $\kappa_\mathbf{k}^2$ {based on the gradient of specific angular momentum $j = R^2 \Omega$}:
\begin{eqnarray}
\kappa_\mathbf{k}^2 = \frac{k_z^2}{k^2}\left(\kappa_R^2 - \frac{k_R}{k_z} \kappa_z^2\right) = \frac{1}{R^3}\frac{k_z^2}{k^2}\left(\partial_R j^2 - \frac{k_R}{k_z} \partial_z j^2\right),
\label{eq:2dKAP}
\end{eqnarray}
and the projected buoyancy frequency\footnote{In the notation of \citet{Urpin2003} this is the term $\omega_g^2$, but we want to stick with the notation of \BVFs.} $N_\mathbf{k}^2$.:
\begin{eqnarray}
 N^2_\mathbf{k}= - \frac{c^2}{\gamma} \frac{(k_R b_z  - k_z b_R) \left(k_R s_z  - k_z s_R  \right)}{k^2} \equiv - \frac{c^2}{\gamma}  \frac{({\bf k} \times {\bf b})\cdot({\bf k}  \times {\bf s})}{k^2},
  \label{eq:BVTK}
\end{eqnarray}
with the logarithmic entropy $\mathbf{s}$ and pressure gradient $\mathbf{b}$. Both $\kappa_\mathbf{k}^2$ and $N^2_\mathbf{k}$ are functions of $R$ and $z$, and depend on the direction of the velocity perturbation, which occurs perpendicular to $\mathbf{k}$.
The indicated analytic growth rates in Figure \ref{Fig:TAUGRATE4Q1} are the fastest rates for either instability, i.e.\ their individual optimum $\mathbf{k}$, see Paper I. As GSF is an instability with no real part in $\omega$ and COS is an overstable oscillation with the real part of $\omega \approx \Omega$, i.e.\ the local epicylclic frequency it is straight forward to separate the COS and GSF growth rates. 

To explain the shape of the growth rates as function of $\tau$ and $z$ it is helpful to analyze the approximate solutions of the dispersion relation.
We find the local GSF growth rates for instantaneous cooling
 \begin{equation}
    \Gamma_\mathrm{GSF}(\tau << \tau_c) = \frac{1}{2}\frac{|\kappa_z^2|}{\Omega^2} \Omega = \frac{|q|}{2} \frac{|z|}{R} \Omega,
\end{equation}
thus height dependent and proportional to $q$.
For GSF as a function of normalized thermal relaxation time $\tau^* = \tau \Omega \gamma$, we found
\begin{equation}
    \Gamma_\mathrm{GSF} = \frac{h^2 q^2}{4} \frac{\gamma}{\gamma -1 }\frac{1}{\gamma \Omega \tau_\mathrm{c,GSF} + \tau^*} \Omega,
   \label{Eq:GAMMA_VSI}
\end{equation}
with the critical cooling time for GSF:
\begin{equation}
    \tau_\mathrm{c,GSF} = h \frac{H}{2 |z|} \frac{|q| }{\gamma -1} \Omega^{-1}.
\end{equation}
which for $z = H/2$ leads to the same result as for the critical time for VSI: $\tau_c$. Thus for long cooling times the growth rate is independent of $z$ and proportional to $q^2$and $h^2$.

Likewise for the COS we derived:
\begin{equation}
   \Gamma_\mathrm{COS} = \frac{h^2 q^2}{8} \frac{\gamma}{\gamma -1 }\frac{\tau^*}{1 + {\tau^*}^2} \Omega,
    \label{Eq:GAMMA_COS}
\end{equation}
which has a maximum for $\tau^*=1$.
For $\tau^*  > 1$, both growth rates GSF and COS attain a constant ratio of $2$ with respect to each other and this result is largely independent of the radial density structure. 
We plot the predicted growth rates as a function of cooling time for a selected range of heights in Fig.\ \ref{Fig:TAUGRATE4Q1} for $q=-1$ and for $q=-0.5$.
We also include the estimates for critical VSI cooling time from \citet{Lin2015} and the growth rates from \citet{Nelson2013} as vertical and horizontal yellow lines, respectively, and find them as good indicators for the asymptotic behavior of the GSF growth rates at large heights for short cooling times.

We find that the GSF growth rates for $\tau^* = 1$ are {approximately} one order of magnitude smaller than the optimal growth rates for GSF at the respective height and COS is therefore 40 times slower. For much longer cooling times like $\tau^* = 10$, both GSF and COS growth times are about two orders of magnitude longer than for the optimal VSI.

We also added our measured values for growth rates from our numerical experiments in the following section to these plots and will discuss them in that section.

The analytic growth rates are predictions for the linear phase of growth. Based on these growth rates we cannot estimate what the saturated level of turbulence may be or, even harder, what the contribution to turbulent angular momentum transport. Both questions, even though very relevant, are beyond the scope of this paper. The fact that we thought disks to be stable with respect to VSI beyond the critical cooling time in our recent simulations \citep{Manger2021} hints at the problem that low growth rates need low numerical dissipation. Thus a careful choice of solving scheme and sufficient spatial resolution is key. A first attempt in establishing this knowledge is part of the following section.

All models for $q=-1$ (and respectively $q=-0.5$) use the same initial density and temperature structure, which also defines the temperature towards which temperature fluctuations are getting relaxed. In the PLUTO code we control the cooling by updating the pressure according to:
\begin{equation}
    P^{t+dt} = \rho T_0 - \left( \rho T_0 - P^{t}\right)e^{-dt / \tau},
\end{equation}
which is stable for an arbitrarily short cooling time.
The Strang splitting method is typically applied in the PLUTO code, i.e.\ one switches the order of hydrodynamic versus cooling operators each timestep. However, this had to be modified to a leap-frog type of splitting, to handle cooling times properly that are smaller than the hydrodynamic step:
\begin{eqnarray}
    P^{t+0.5 dt} &=& \rho T_0 - \left(\rho T_0 - P^{t}\right)e^{-0.5 dt / \tau},\\
    P^* &=& f_\mathrm{hydro}(P^{t+0.5 dt}),\\
    P^{t+dt} &=& \rho T_0 - \left(\rho T_0 - P^*\right)e^{-0.5 dt / \tau}.
\end{eqnarray}
The cooling time is varied from $\tau^* = 10^{-5}$ to $\tau^* = \infty$ (no cooling) and the initial perturbation is $10^{-4} $ of density, but the pressure remains unchanged, i.e.\ we perturb the temperature inversely to the density.

\begin{figure*}
\gridline{\fig{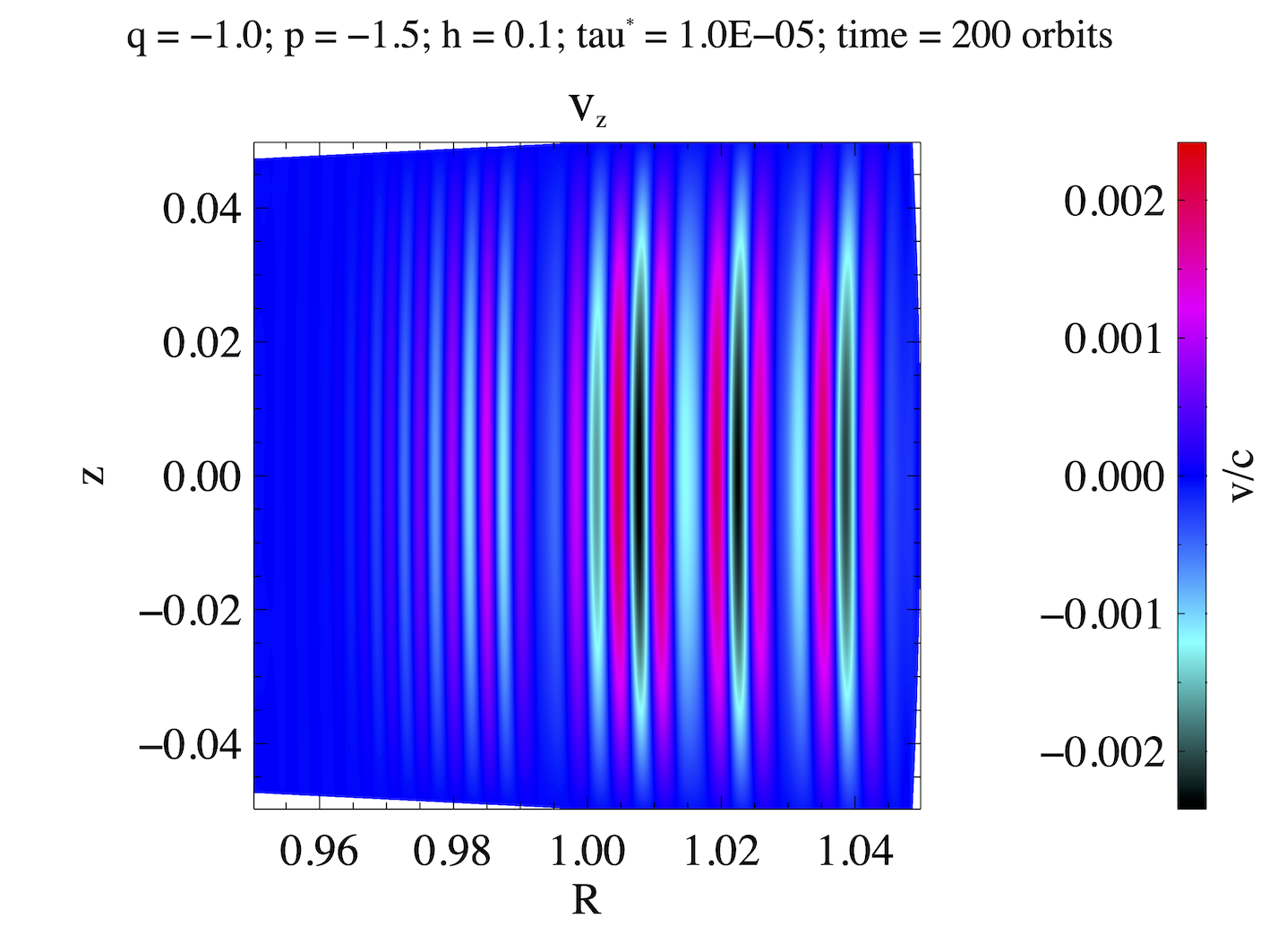}{0.5\linewidth}{(a): $z_0 = 0; t = 200 \,\mathrm{ Orbits}$}\fig{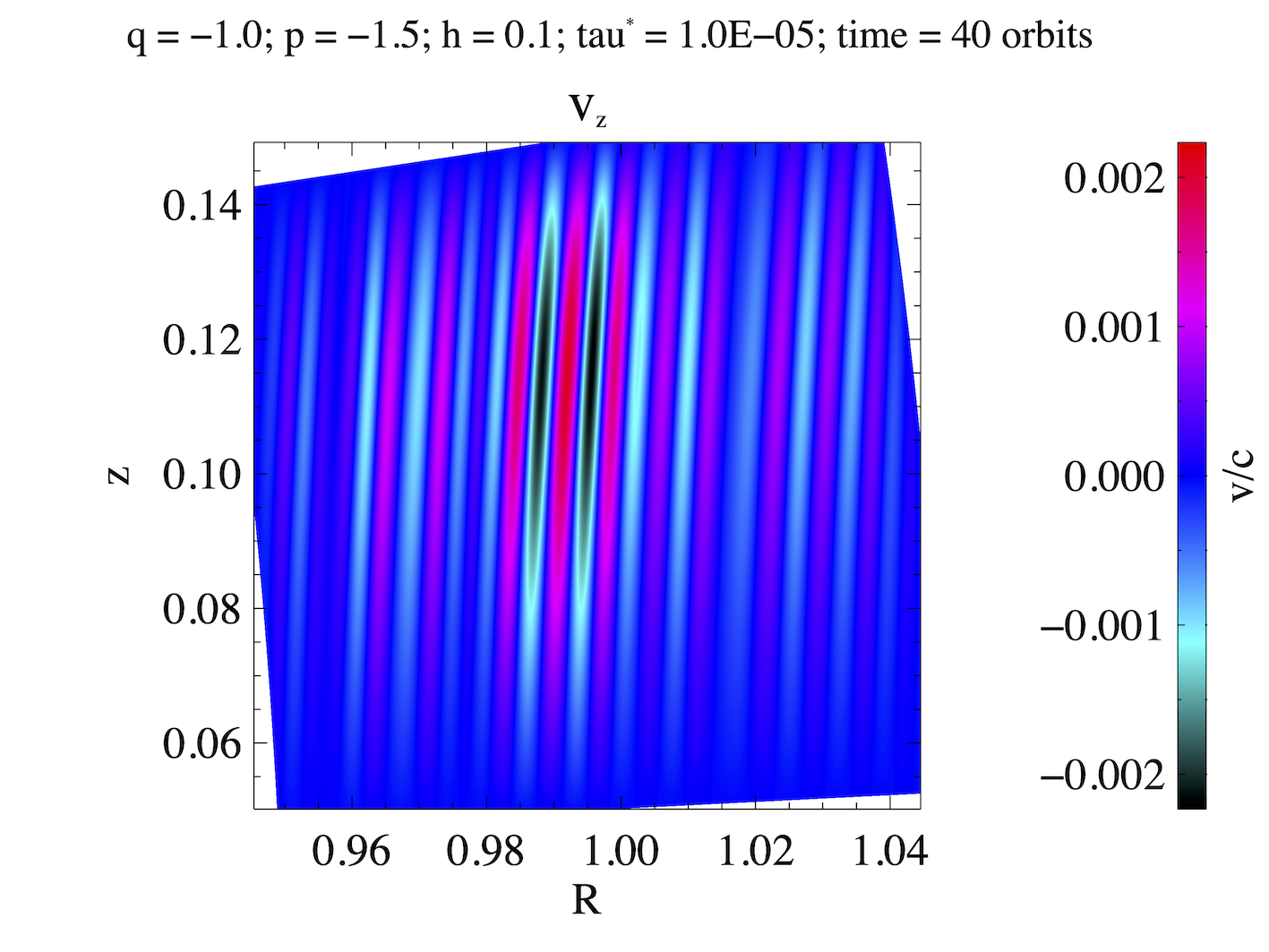}{0.5\linewidth}{(b): $z_0 = H; t = 40 \,\mathrm{ Orbits}$}}
  \caption{Classical GSF for short cooling times: {${\tau^*} = 10^{-5}$} Simulation snap shots for $q = -1, p = -1.5, $. (a): for the midplane ($z = [-0.5H, 0.5H]$) and (b): for the atmosphere ($z = [0.5H, 1.5H]$). We plot the 
vertical velocities ($v_z$) indicated with a red (positive = upward) and black/cyan (negative = downward) color scheme. Same simulations as in Fig.\ \ref{Fig:VMAXQ1T0}.
    \label{Fig:VSIQ1T0}}
\end{figure*}
\begin{figure*}
\gridline{\fig{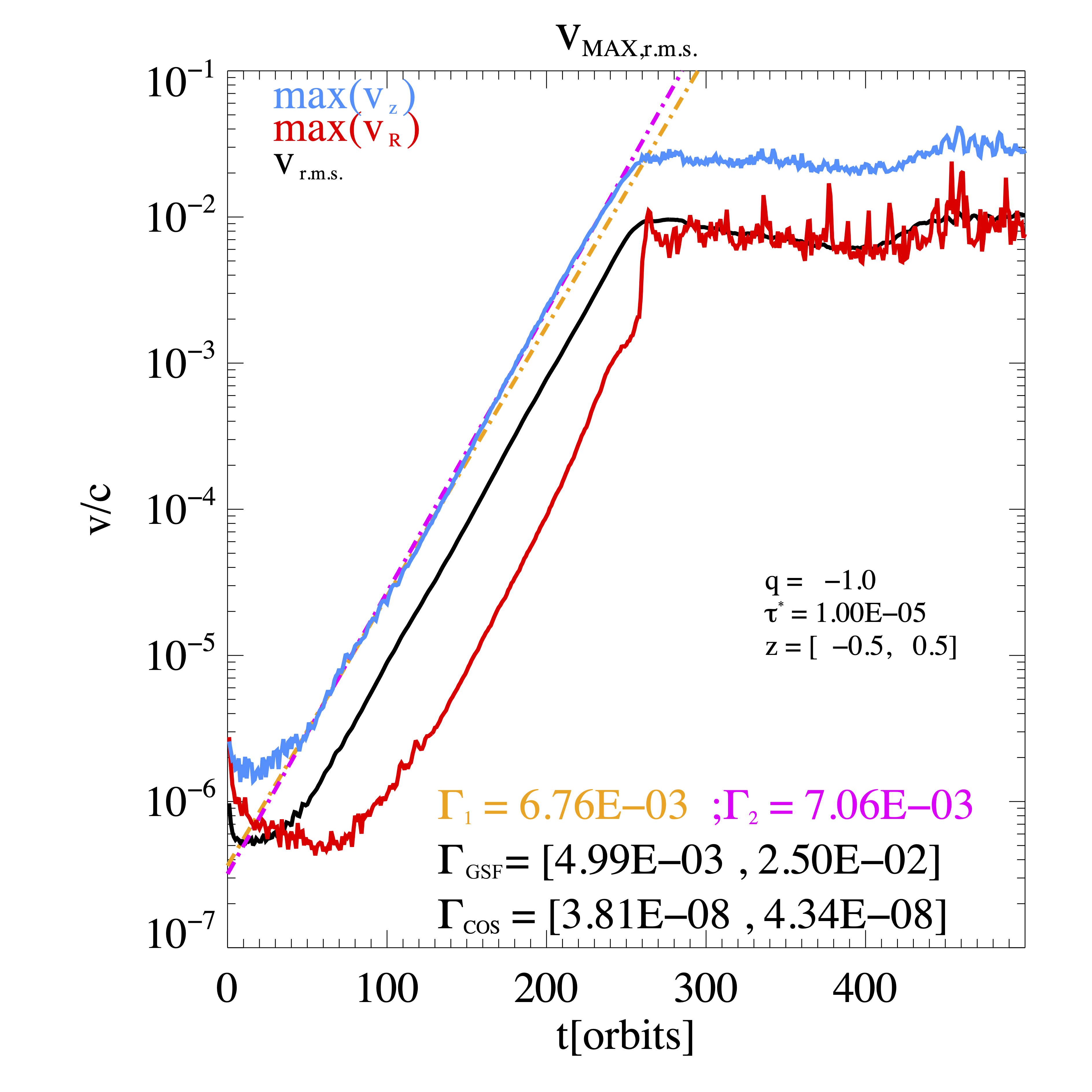}{0.5\textwidth}{(a): $z_0 = 0$}\fig{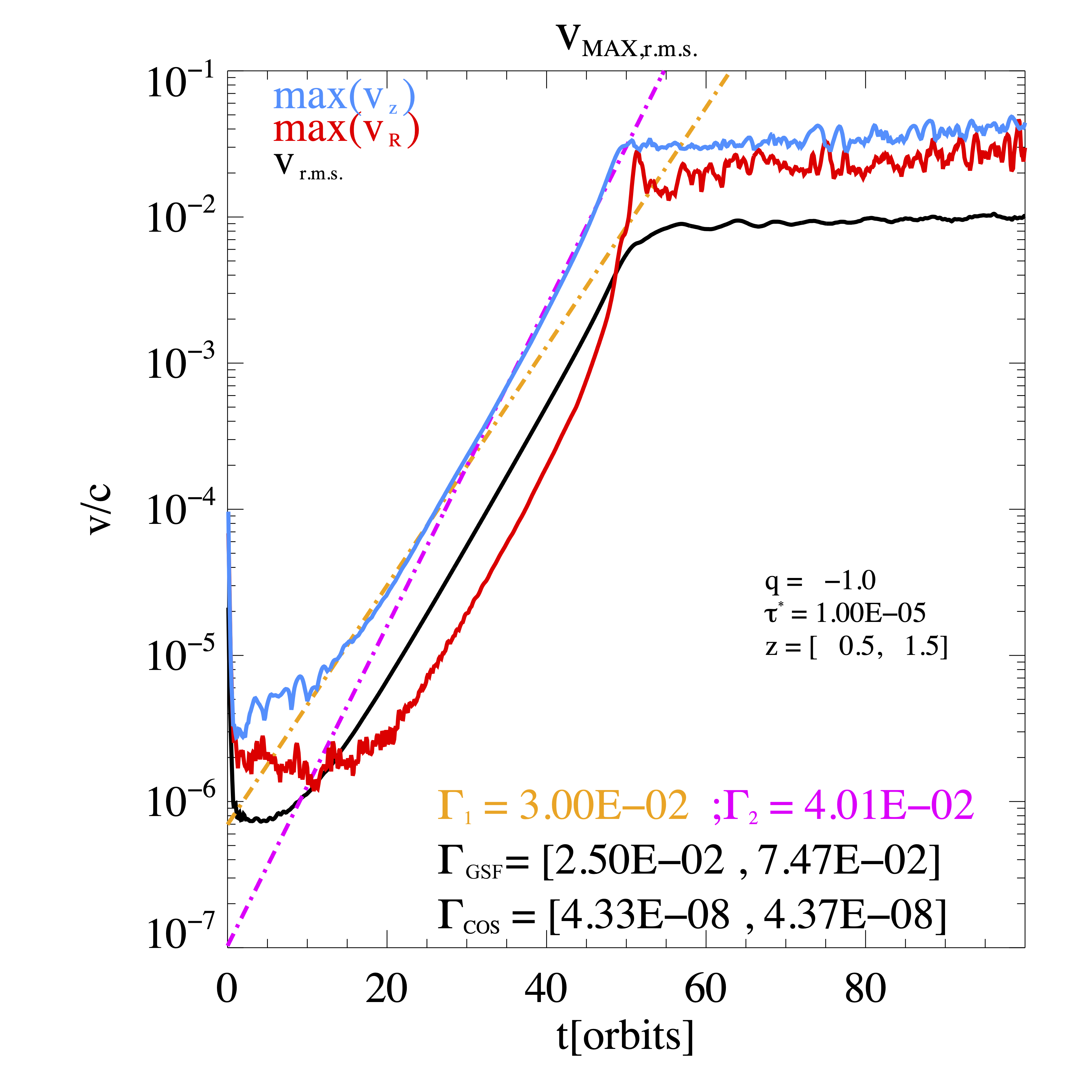}{0.5\textwidth}{(b): $z_0 = H$}}
  \caption{Evolution of r.m.s.\ and maximal velocities for $q = -1, p = -1.5, \tau^* = 10^{-5}$. (a): for the midplane ($z = [-0.5H, 0.5H]$) and (b): for the atmosphere ($z = [0.5H, 1.5H]$), in units of the speed of sound and as function of time in units of orbits. We show the r.m.s. velocities (in the $R$-$z$ plane) as solid black line. The largest radial velocity is indicated with the red line and the largest vertical velocity with the blue line. The magenta and yellow lines are the fitted growth rates $\Gamma_1$ and $\Gamma_2$ for two different time intervals given in plot with the analytic growth rate range for the simulated domain.
    \label{Fig:VMAXQ1T0}}
\end{figure*}

It is not trivial to study the linear evolution phase for an instability with a Godunov solver aimed to be stable and accurate in the presence of shocks. One would actually wish to use a low Mach-number code \citep{Almgren2006,Edelmann2021}, which is a topic for future projects, or a high-order finite difference method such as the Pencil Code \citep{Lyra2011,Lyra2014}. In addition, a spectral code with a Boussinesq ansatz as in the code SNOOPY, has been successfully used for VSI simulations \citep{Latter2022} as well as for unstratified COS simulations \citep{Teed2021}. Unfortunately, in quasi-incompressible local Boussinesq simulations the ability of sound waves to carry angular momentum \citep{Heinemann2009} is suppressed. Thus one needs global compressible simulations to study transport processes as well as the possible formation of particle traps like zonal flows and vortices. For the moment, we use a code that we know can handle global VSI simulations \citep{Manger2021,Pfeil2021}. If our code cannot reproduce the linear growth rates, we would also not trust the results of a three dimensional simulation.

With the PLUTO 4.3 code \citep{Mignone2007} we found that high order time integration (Runge Kutta 3) and space interpolation (WENO3) in combination with an accurate Riemann solver ({ {in our case a Roe solver}) is essential for converging results at reasonable resolution. Parabolic interpolation of 5th order in space is slightly less suitable\footnote{Recently we switched to PLUTO 4.4 and now can use the WENOZ scheme in combination with HLLC for fast and accurate reproduction of the linear growth phase.}. We tried both conservation of total energy and entropy conservation for the energy equation, with little difference in the linear phase. Thus our simulations use the total energy scheme of PLUTO, which is numerically slightly cheaper than the entropy conserving scheme.

In Fig.\ \ref{Fig:TAUGRATE4Q1}, we plot the measured growth rates for $q=-1$ for the midplane centered boxes (models A: crosses $+$) and $z_0=H$ centered boxes (models B: $X$) along with the analytic predictions for GSF (blue lines) and COS (red lines) for various heights. The {dotted} line {indicates the separation} between the two sets of simulations, i.e.\ growth rate for $z = 0.5 H$. Thus, we find a good reproduction of the predicted growth rate. For short cooling times, the GSF modes clearly must be the drivers of growth. For longer cooling times, the simulations in the higher atmosphere clearly show the right amount of decrease in growth as expected for GSF, yet for the midplane boxes the measured growth rates could be produced by either COS or GSF. We therefore inspect a trio of models with a range of cooling times $\tau^* = 10^{-5}, \, 1, \, 10$ more closely, which will also show how we obtained the plotted growth rates for Fig.\ \ref{Fig:TAUGRATE4Q1}.

\subsubsection{Models: $\tau^* = 10^{-5}$}
Models with extremely short cooling timescales are the closest to the classical simulations of VSI \citep{Nelson+2016,Stoll2014,Manger2018,Manger2020,Manger2021,Pfeil2021}. In Fig.\ \ref{Fig:VSIQ1T0}, we show snapshots of the vertical velocities and temperature perturbations during the linear growth phase. One clearly recognizes the radially alternating vertical motion of the gas driven by the VSI/GSF.
With careful inspection one finds that the direction of the vertical motion is neither purely vertical nor along contours of constant angular momentum, but half-way between both.
This elucidates the cause of the linear GSF modes, as a flow of higher angular momentum material upward and more importantly outward into a region of lower angular momentum, just as in the classical Rayleigh criterion for rotational stability. This is the fundamental cause of the GSF and thus VSI.

In Fig.\ \ref{Fig:VMAXQ1T0}, we show the time evolution of the velocities. We plot for both models the overall r.m.s.\ velocities of the radial and vertical components in order to measure the growth rates the largest velocities. We plot vertical velocities in blue and radial velocities in red; once more one recognizes that for both the midplane and the atmosphere models the GSF dominates, even at the midplane with the smaller expected growth rate. The saturation level of the r.m.s.\ speed in both cases is very similar, despite the difference in growth rate.
Interestingly, we observe a different saturation effect for GSF modes than the effects of saturation discussed for global VSI simulations by \citet{Latter2018} and \citet{Cui2022}, probably due to the local nature of our simulations. In fact, after the linear growth phase, clockwise rotating eddies emerge, very similar to the ones we show in the following paragraph, where we will discuss them some more.

\begin{figure}
\includegraphics[width=0.9\linewidth]{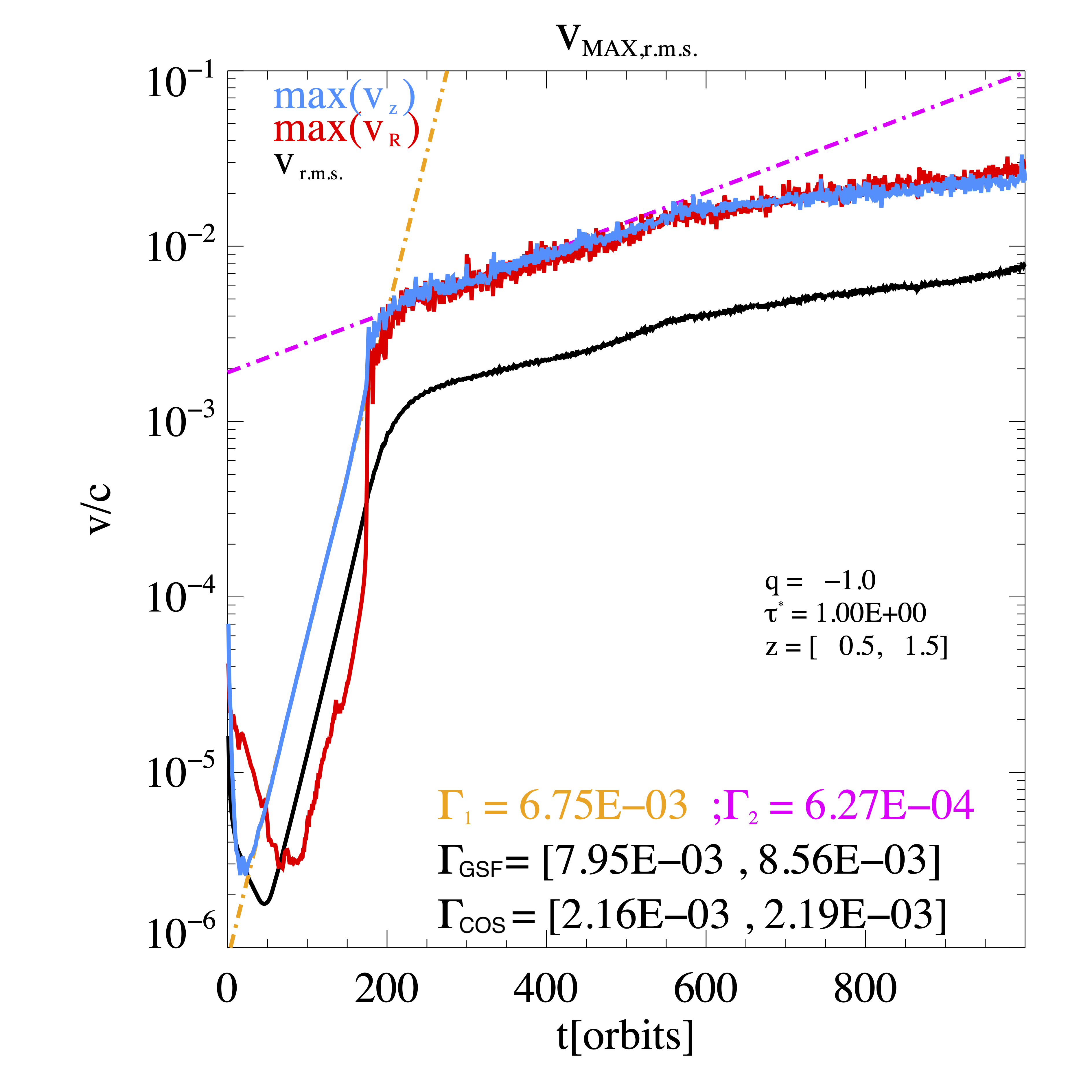}
  \caption{Evolution of r.m.s.\ and maximal velocities for $q = -1, p = -1.5, \tau^* = 1$ in  the disk atmosphere ($z = [0.5H, 1.5H]$) in units of the speed of sound and as function of time in units of orbits. We show the r.m.s. velocities (in the $R$-$z$ plane) as a solid black line. The largest radial velocity is plotted with a red line and the largest vertical velocity with a blue line. The yellow and magenta dashed dotted lines are the fitted growth rates $\Gamma_1$ and $\Gamma_2$ given in plot, along with the analytic growth rate range for the simulated domain.
    \label{Fig:VMAXQ1T1}}
\end{figure}
\begin{figure}
\gridline{\fig{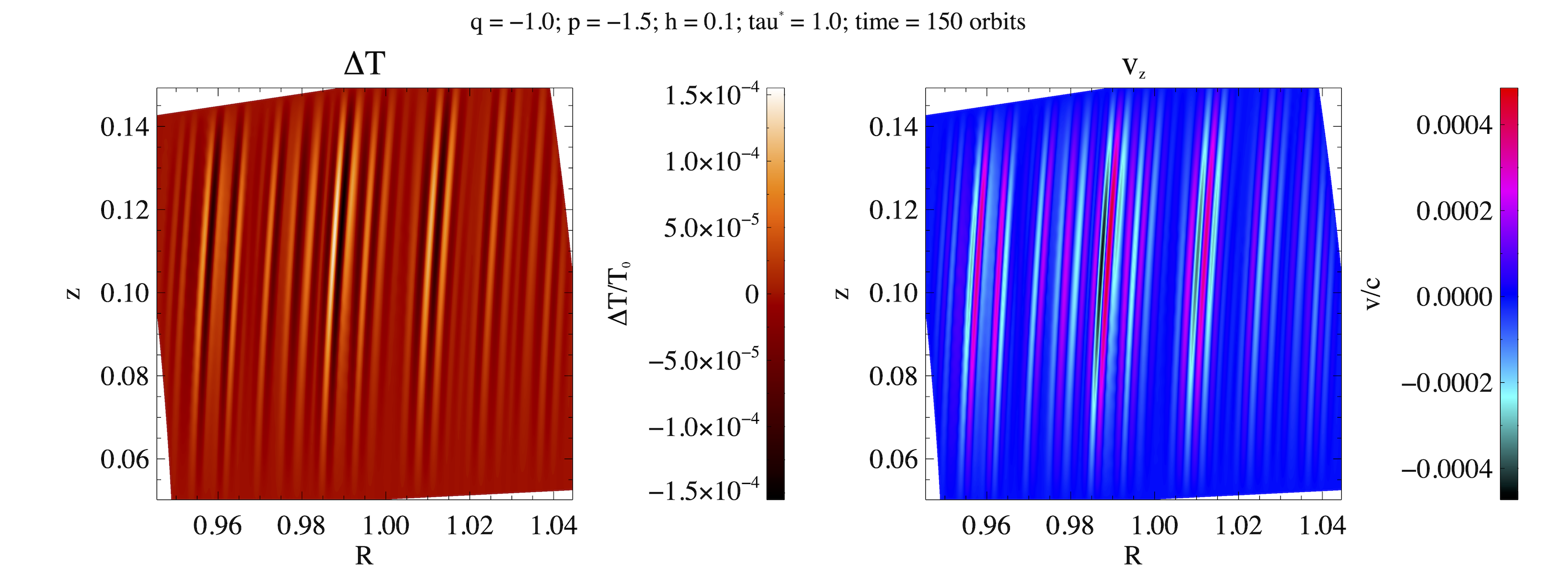}{1.0\linewidth}{(a): $t = 150 \,\mathrm{ orbits}$}
}
\gridline{\fig{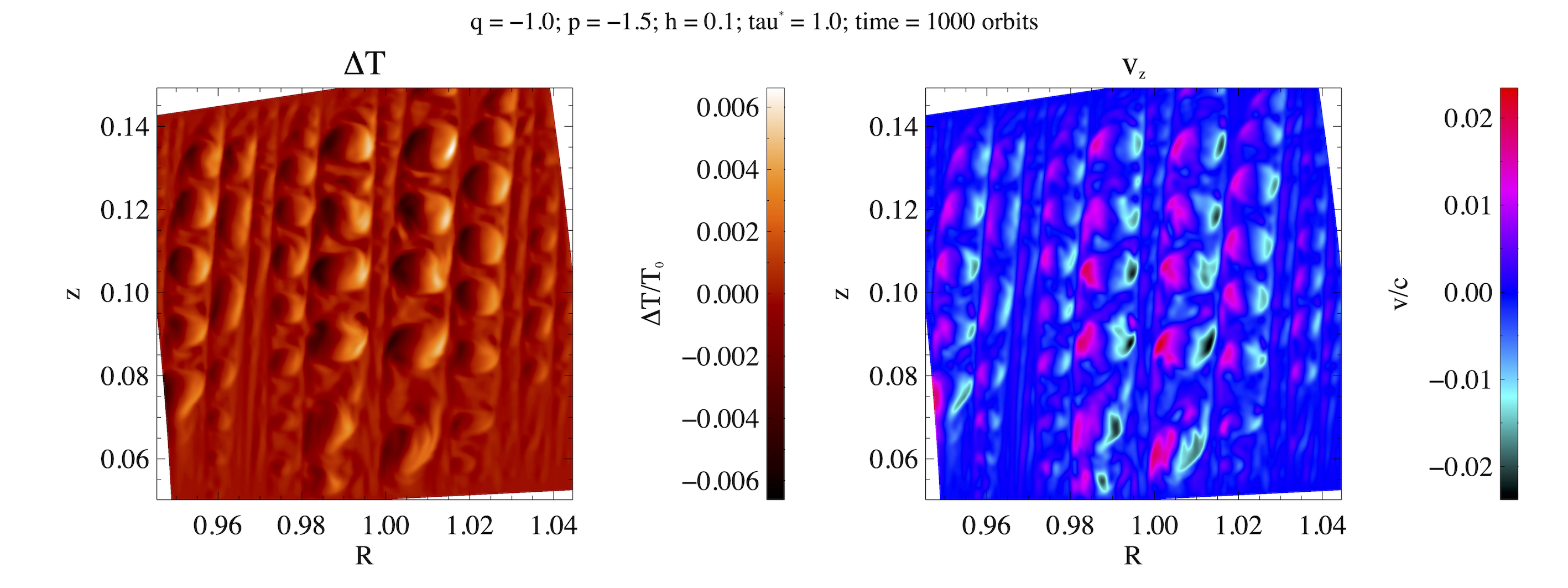}{1.0\linewidth}{(b): $t = 1000 \,\mathrm{ orbits}$}
}
  \caption{Simulation snapshots for $q = -1, p = -1.5, \tau^* =1 $. An early snapshot showing the linear development of GSF modes (a) and a later snapshot with showing the non-linear pattern of circulation regions (b), both in the atmosphere ($z = [0.5H, 1.5H]$).
  We plot the deviation of local temperature $\Delta T = T'/T_0$ and
   vertical velocities in the $R$-$z$ plane. The further development of this simulation can be found in Fig.\ \ref{Fig:VSIQ1T1B}.
    \label{Fig:VSIQ1T1}}
\end{figure}

\begin{figure*}
\gridline{\fig{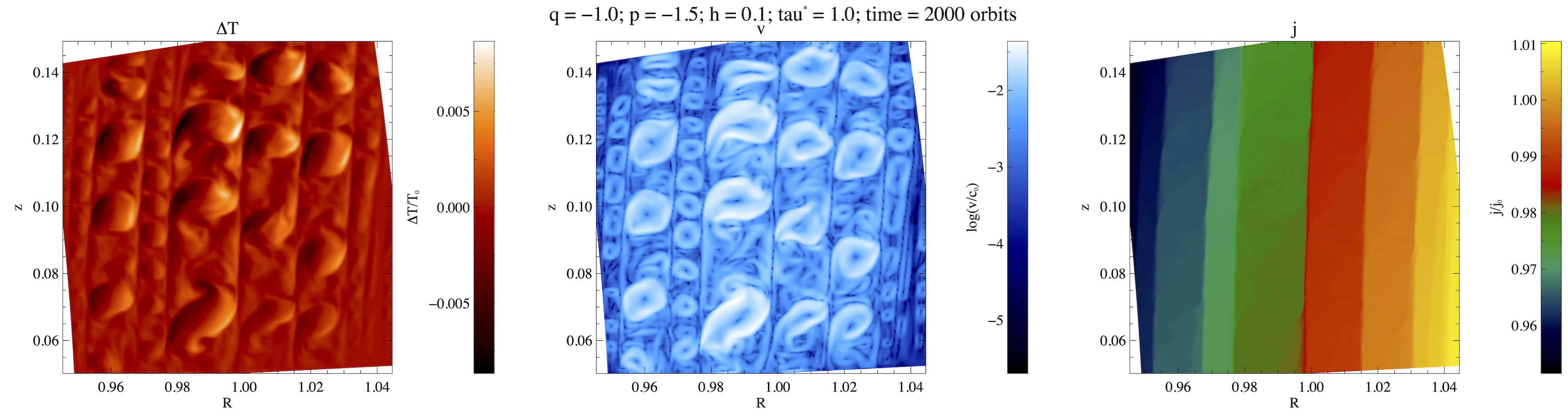}{\textwidth}{(a): $t = 2000 \,\mathrm{ orbits}$}
}
\gridline{\fig{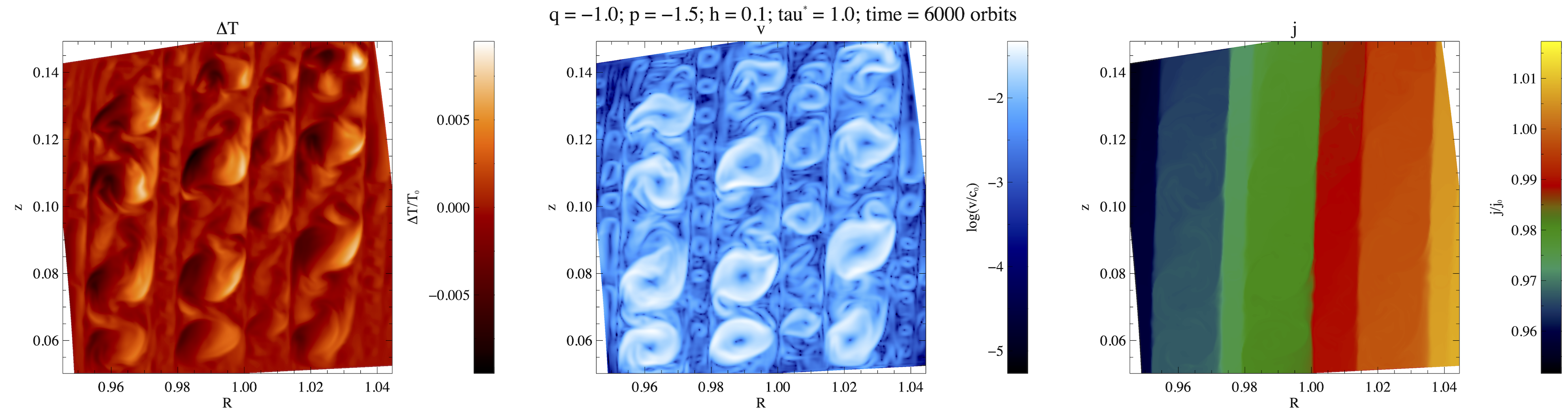}{\textwidth}{(b): $t = 6000 \,\mathrm{ orbits}$}
}
  \caption{Simulation snapshots for $q = -1, p = -1.5, \tau^* =1 $ showing the non-linear pattern of circulation regions (b), both in the atmosphere ($z = [0.5H, 1.5H]$) sitting in bands of locally constant angular momentum $j = \Omega R^2$. We plot the deviation of local temperature $\Delta T = T'/T_0$,
   absolute velocities in the $R$-$z$ plane indicate pseudo-streamlines, the darkest regions correspond to the largest velocities and the white regions to zero velocity. 
    \label{Fig:VSIQ1T1B}}
\end{figure*}

\subsubsection{Models: $\tau^* = 1$}
For a cooling time $\tau^* = 1$, the COS modes are expected to reach their fastest growth and the GSF modes grow significantly slower than in the $\tau = 10^{-5}$ case, yet still about four times faster than the COS (see Fig.\ \ref{Fig:VMAXQ1T1}). In Fig.\ \ref{Fig:VSIQ1T1}, we see that for the $z_0 = H$ case, one still recognizes the typical GSF modes during the linear growth phase and Fig.\ \ref{Fig:VMAXQ1T1} confirms that the measured growth rates coincide with the predicted GSF growth rates for the first 200 orbits. After the linear GSF growth comes to a halt, the system slowly continues to evolve, still growing, but at a lower rate. At this stage, the developing pattern is no longer dominated by vertical motion, but forms small loops in the $R$-$z$ plane as can be seen in Fig.\ \ref{Fig:VSIQ1T1}. The loops are initially located in the bands of constant angular momentum, created by the linear phase of the GSF and over time create more extended radial regions of constant angular momentum (See \href{https://youtu.be/YeB_fOlLwsE}{Movie 1a}).

We found that these eddies are not driven by convection, as we measured no down gradient corresponding transport of entropy, but instead the opposite. They mix entropy downward to the midplane. Note that they also appear in the $\tau^* = 10^{-5}$ simulation, where a short cooling time prevents any convection. The eddies can also not be driven by vertical shear, as they sit in bands of constant angular momentum, where there is no vertical shear. In a way the GSF has produced a state in which it cannot operate anymore, which defined the end of its growth.

 \begin{figure}
\includegraphics[width=1.0\linewidth]{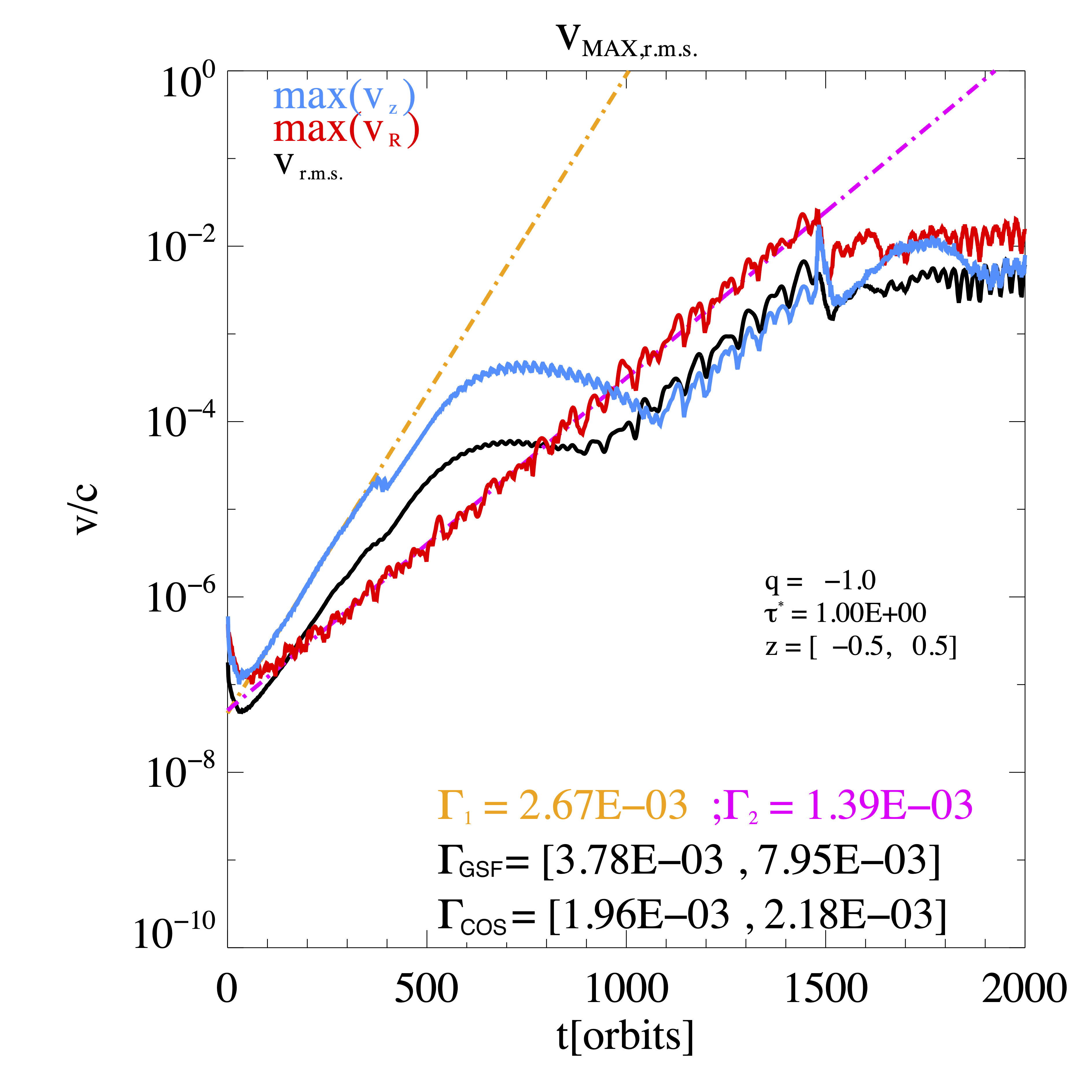}
  \caption{Evolution of r.m.s.\ and maximal velocities for $q = -1, p = -1.5, \tau^* =1$ in the disk midplane ($z = [-0.5H, 0.5H]$) in units of the speed of sound and as function of time in units of orbits. We show the r.m.s. velocities (in the $R$-$z$ plane) as a solid black line. The largest radial velocity is plotted with a red line and the largest vertical velocity with a blue line. The yellow and magenta dashed dotted lines are the fitted growth rates $\Gamma_1$ and $\Gamma_2$ given in plot, along with the analytic growth rate range for the simulated domain.
    \label{Fig:VMAXQ1T1B}}
\end{figure}

\begin{figure}
\gridline{\fig{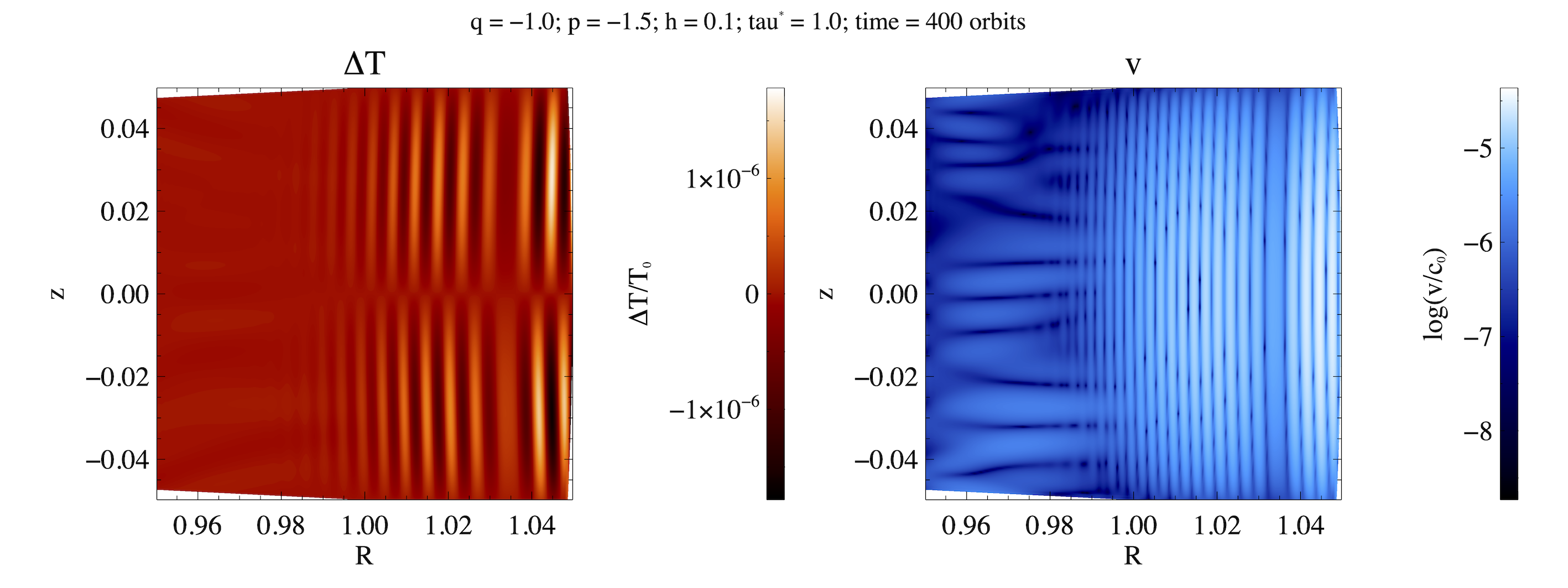}{\linewidth}{(c): $z_0 = 0; t = 400 \,\mathrm{ orbits}$}
}
\gridline{\fig{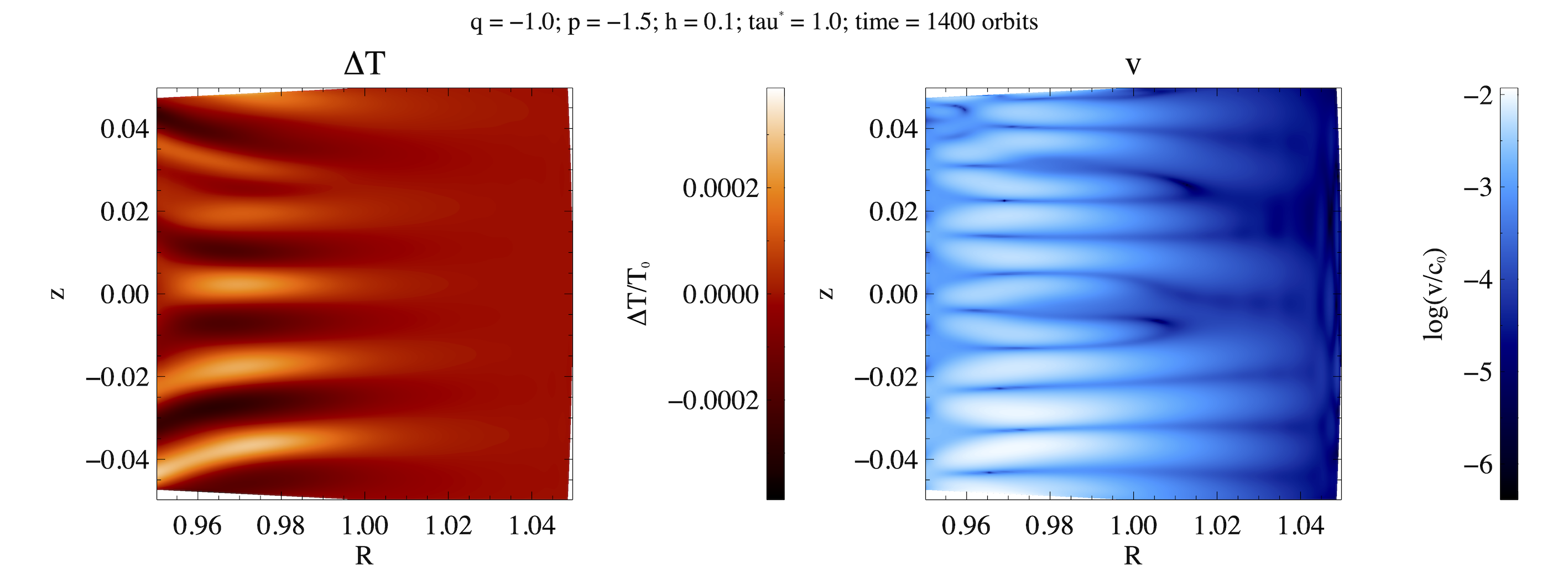}{\linewidth}{(d): $z_0 = 0; t = 1400 \,\mathrm{ orbits}$}
}
\gridline{\fig{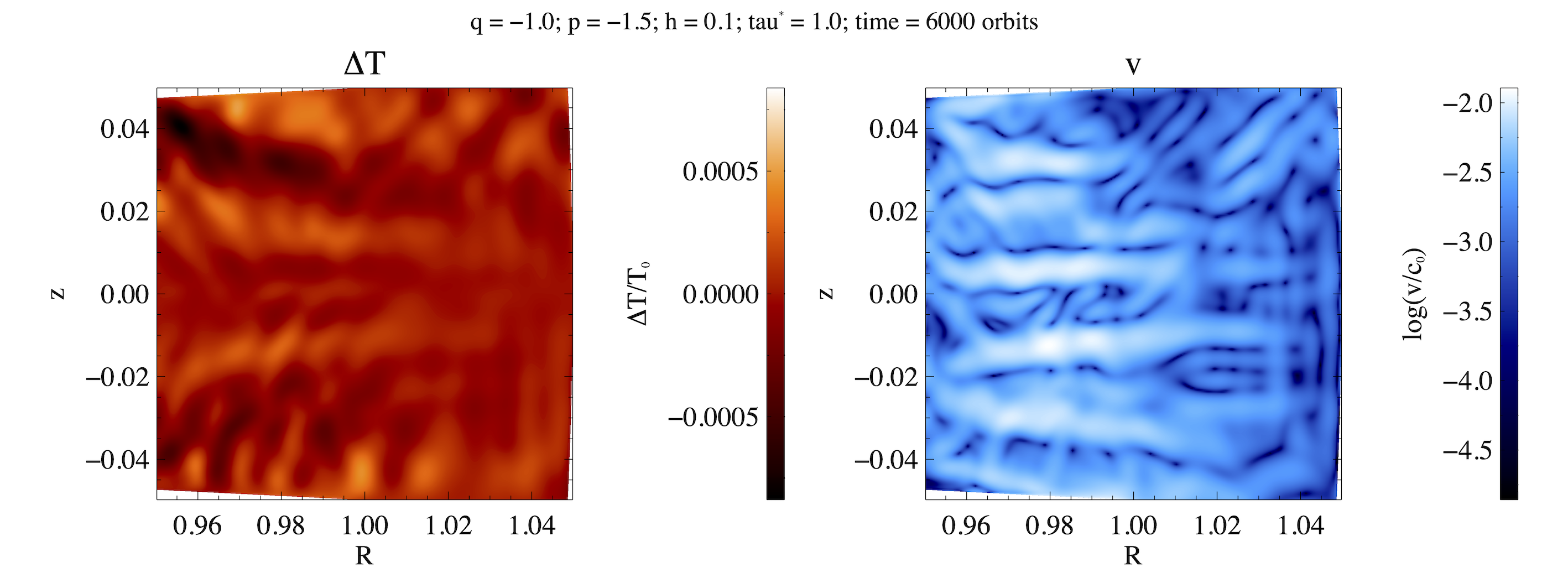}{\linewidth}{(e): $z_0 = 0; t = 6000 \,\mathrm{ orbits}$}
}
  \caption{Simulation snapshots for $q = -1, p = -1.5, \tau^* =1 $ (c), (d) and (e) for the midplane ($z = [-0.5H, 0.5H]$). Continuation of Fig.\ \ref{Fig:VSIQ1T1}. See also Fig. \ref{Fig:VMAXQ1T1}.
    \label{Fig:VSIQ1T1L}}
\end{figure}

The situation at the midplane is also very interesting. Initially, vertical GSF-like modes dominate, and later radial COS modes take over, as can be seen in the snapshots (Fig.\ \ref{Fig:VSIQ1T1}) as well as in the growth rates (Fig.\ \ref{Fig:VMAXQ1T1}). Initially, the largest velocities are the vertical motions (blue curve) and later from about 1000 orbits onward, radial oscillations (red curve) are stronger. In fact, COS modes grow right from the start, and we have a nice superposition of COS and GSF modes, which in their linear stage do not affect each other.

After about 1700 orbits, the radial oscillations of the epicycles are so strong that the vertical shear between an outward moving and inward moving band becomes unstable to vertical shear modes. Briefly, the vertical velocities dominate over the radial velocities, damping the radial velocities, which then start to grow again. This can best be seen in a \href{https://youtu.be/02TnKk5rNv0}{movie} created from this simulation. 
\citet{Lyra2014} and \citet{Teed2021} already discuss various phases of growth for unstratified COS simulations and the second phase was attributed to a Kelvin-Helmholtz instability (KHI) acting as a parasitic effect on the channel modes (radial epicyclic COS oscillations). In our stratified simulations, we argue that saturation of COS modes occurs when the channel modes which conserve angular momentum reach an amplitude at which the oscillating vertical shear between inward and outward moving sheets of gas becomes unstable as a variant of the GSF itself. More precisely, as soon as the GSF growth time, based on the amplitude of the oscillation of vertical shear, is shorter than the oscillation period, we get a small outburst of GSF that removes part of the vertical shear and thus damps the radial COS modes. After that the COS grows again, eventually leading to the next GSF eruption (see \href{https://youtu.be/02TnKk5rNv0}{Movie 1b}).
Note that this GSF is not product of the baroclinic state of the disk, but a secondary instability once the COS modes have an appropriate amplitude.

\subsubsection{Models: $\tau^* = 10$}
For even longer cooling times, we still find the evolution of GSF modes in the atmosphere and clear COS modes in the midplane. In the latter case, the radial modes dominate throughout the run. One also finds the intermittent outbreak of GSF modes, once COS reaches a certain amplitude (see Fig.\ \ref{Fig:VSIQ1T6}). 

\begin{figure}
\gridline{\fig{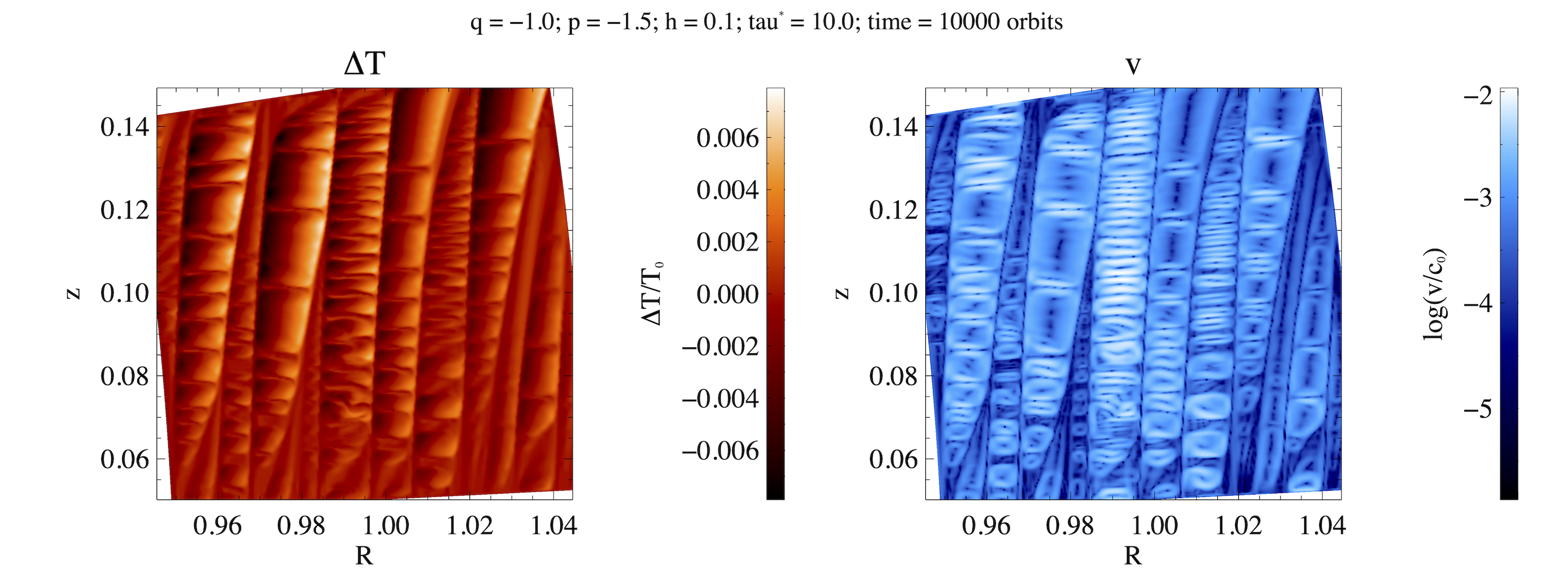}{\linewidth}{(a): $z_0 = H; t = 10000 \,\mathrm{ orbits}$}
}
\gridline{\fig{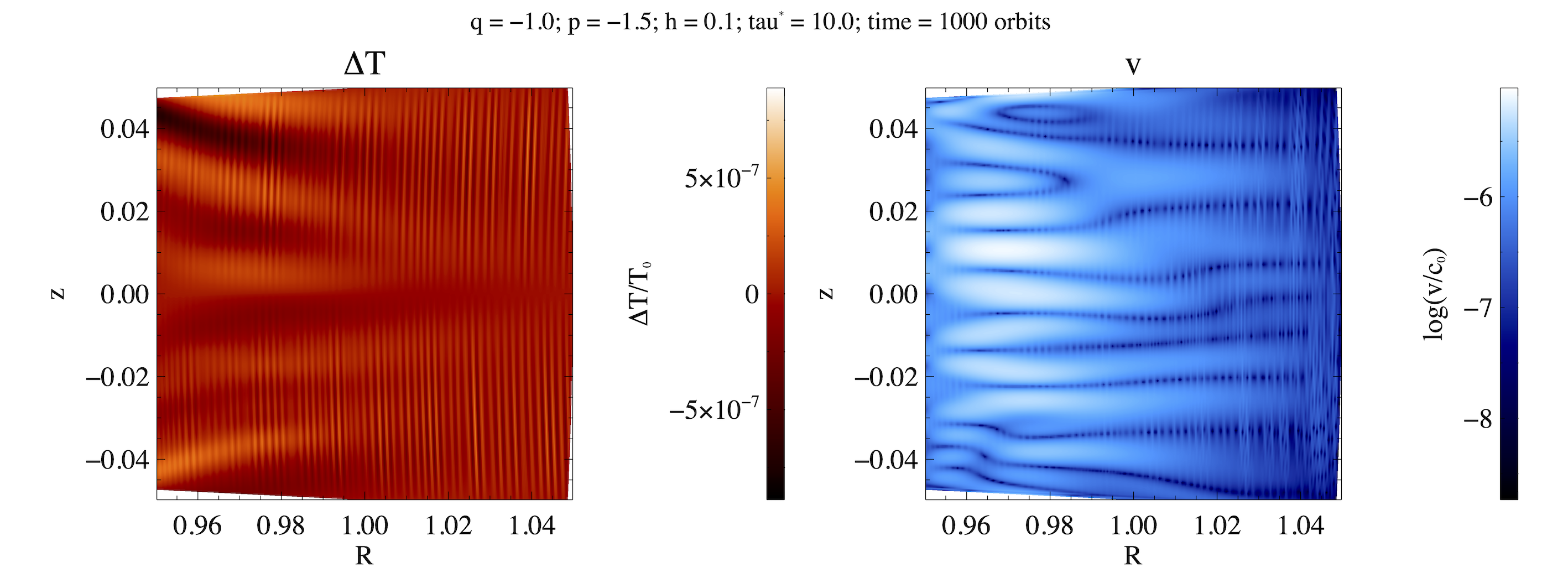}{\linewidth}{(b): $z_0 = 0; t = 4821 \,\mathrm{ orbits}$}
}
  \caption{Simulation snapshots for $q = -1, p = -1.5, \tau^* =10 $, (a) for the atmosphere ($z = [0.5H, 1.5H]$) and (b), for the midplane ($z = [-0.5H, 0.5H]$). We plot the 
  deviation of local specific entropy $K = P \rho^{-\gamma}$ with $\Delta K = K'/K0$ in units of the background entropy. Lighter indicates more entropy, darker less entropy. Absolute velocities in the $R$-$z$ plane indicate pseudo-streamlines, the darkest regions correspond to the largest velocities and the white regions to zero velocity, overall normalized to the velocities at the given time.
    \label{Fig:VSIQ1T6}}
\end{figure}

\subsubsection{Models: $q=-0.5$}
We did the same exercise with simulations using a shallower temperature gradient of $q=-0.5$ and our measurements of linear growth rates can be found in Figure \ref{Fig:TAUGRATE4Q1}. For the $z_0=H$ runs we found good agreement between the predicted GSF growth rates and the measured values, but not as good as in the $q=-1$ case. We explain this discrepancy by numerical dissipation, which damps the growth of modes. Therefore we added runs with double the resolution (512 cells per scale height) and measured growth rates much closer to the analytic prediction.

For the models in the midplane $z_0 = 0$, the measured growth rates are much lower than predicted, especially for longer cooling times. The missing crosses in Figure \ref{Fig:TAUGRATE4Q1} indicate simulations for which we did not find a linear growth-phase, despite growing velocities and thus could not measure a proper growth rate.

One possible continuation these studies, would be to increase the resolution, which might find stronger growth rates and extended linear periods of growth, but might require a switch to a much less dissipative scheme. Note that by increasing the resolution, we also end up with a smaller time step, which means that our sub-sonic motions do not benefit as much as an incompressible scheme by the increased resolution. An incompressible scheme like Snoopy \citep{Lesur2005} or possibly a low-Mach number scheme \citep{Almgren2006} should be able to clarify this in the future.

Nevertheless, based on our numerical experiments, we argue that the predicted growth rates for our dispersion relation can actually be recovered in non-linear simulations.
We can identify GSF modes and COS modes in the linear evolution of vertically vs.\ radially dominant velocity perturbations. 
Even more importantly, we can confirm that GSF does not die out at longer critical cooling times, but simply takes a little longer to grow.
But clearly we have shown that one cannot construct a stable atmosphere for a protoplanetary disk with a radial temperature gradient that has a non-vanishing thermal relaxation, i.e.\ any realistic protoplanetary disk.

So far we identified GSF and COS modes in our setups by checking growth rates and dominant directions of velocity. Dominant vertical motions should correspond to GSF modes and dominant radial oscillations to COS modes. But this only holds for {the first hundred orbits} of the linear growth phase. For instance, in the model depicted in Fig.\ \ref{Fig:VSIQ1T1} and \ref{Fig:VSIQ1T1B} with $q=-1$ around $z_0 = H$, we find the end of the linear growth at about 600 orbits (see Fig.\ \ref{Fig:VMAXQ1T1}), after which radial and vertical velocities are similar in amplitude and the growth rates are still positive at a level of $\Gamma = 4 \times 10^{-4}$. The question is whether this continued growth, which is distinguished by the development of little whirls inside the bands of roughly constant angular momentum, is originally created by the VSI and GSF or by the COS? Inspecting the expected growth rates for COS, which are five times stronger than the measured ones, and which can be reproduced in the linear phase of midplane evolution, makes it hard to explain the discrepancy by the dissipative scheme of the PLUTO simulations.

It could also be that this is actually already a non-linear growth phase or at least a new linear growth phase, in which the background state has significantly changed and needs a new analysis.
Before one starts such an attempt, we can analyze the driving force of this swirl amplification. We first measured the net entropy transport along the stream lines of the swirl in order to interpret the swirl as a convection cell that converts buoyancy into motion.
But this analysis of the flow does not support the idea that the swirls are convection cells driven by entropy transport. In fact, we found entropy to be mostly mixed downward to the midplane, as expected.

So to see if these modes depend on the vertical shear or the entropy gradient, we choose to either switch off GSF or suppress COS and see what happens in such a case. We have already argued that in a real disk this is not possible, so we apply a slightly artificial disk simulation in the following section.

\subsection{Separating the modes}
In a real disk, it is impossible to separate GSF and COS for larger cooling times, because for all possible configurations of our atmospheres we found the growth rates of GSF to dominate. Furthermore, for cooling times suitable for COS, the growthrates GSF and COS scale {similarly with respect to} the disk structure, as shown in Equations \eqref{Eq:GAMMA_COS} and \eqref{Eq:GAMMA_VSI}. The cause for both instabilities is the magnitude of the baroclinic term in the vertical radial structure of the disk, i.e.\ the vertical gradient of {angular momentum} being proportional to the {$\phi$ component of the cross product} of density and pressure gradients:
\begin{equation}
\kappa_z^2 = -\frac{1}{\rho^2} \left(\nabla \rho \times  \nabla P\right)_\phi,
\label{eq:radbalance2}
\end{equation}
However, we can use a trick in our numerical simulations. We can modify the applied gravitational forces in a way such that they are no longer conservative (derived from a potential), but instead they either balance the radial buoyancy to remove vertical shear, or they introduce vertical shear for disks without radial temperature gradient.

To simplify things, we first define vertical gravity in the $z \ll R$ limit
\begin{equation}
    g_z = - M G \frac{z}{R^3} = - \Omega^2 z,
\end{equation}
instead of full gravity with the spherical radius $r^2 = z^2 + R^2$. The associated density structure is then the classical Gaussian shape
\begin{equation}
    \rho = \rho_0 e^{-\frac{z^2}{2 H^2}}.
\end{equation}
The full radial component of gravitational acceleration in this $z \ll R$ approximation would be
\begin{equation}
    g_R = -  \frac{M G}{R^2},
\end{equation}
but in order to balance the thermal wind equation (see Eq.\ \ref{eq:radbalance2}) and enforce $\kappa_z^2 = 0$, even when $q \ne 0$, we use a radial acceleration that changes slightly with height
\begin{equation}
    g_R = - \frac{M G}{R^2}\left(1 - \frac{3 + q}{2}\frac{z^2}{R^2} \right).
\end{equation}
This corresponds to a tiny $1\%$ modification of gravity at $z = H = 0.1 R$ for $q=-1$.
The resulting equilibrium rotation profile for the initial condition is then
\begin{equation}
    \Omega^2 = \frac{M G}{R^3}\sqrt{1 + (p+q) \frac{H^2}{R^2}}.
\end{equation}
In such a setup, the disk will allow for the convective modes (COS) but not for the vertical GSF shear modes.

If, on the other hand, we wish to suppress the convective modes, we can set $q=0$ for the actual density and temperature structure of the disk. In contrast to the above strategy, we modify the radial gravity to produce a $\kappa_z^2$ appropriate for a hypothetical temperature gradient $q^*$:
\begin{equation}
    g_R = - \frac{M G}{R^2}\left(1 - \frac{3 - q^*}{2}\frac{z^2}{R^2} \right),
\end{equation}
which implies that
\begin{equation}
    \Omega^2 = \frac{M G}{R^3}\sqrt{1 + q^* \frac{z^2}{2 H^2} + p \frac{H^2}{R^2}}.
\end{equation}
and radial buoyancy, and therefore COS, is eliminated but not the GSF.

We perform these simulations in cylindrical ($R,z$) coordinates for the same parameters $q=-1$, $\tau^*=1$ and $z_0=H$ as for the spherical model, but use an even smaller simulation domain to properly resolve the instabilities without a burdensome numerical cost. We also desire to minimize the effect {of} using non-conservative gravity, because as in M.C.\ Escher's famous infinite staircase, it could be possible to circulate in a closed {streamline} in our simulation domain and continually release potential energy.
Using a simulation with instantaneous cooling but suppressed GSF modes, we tested that this effect does not lead to an artificial instability in our simulations.

The radial and vertical domain centered around a point at $z_0 = H$ is now only $0.2 H$ wide thus $0.99 < R < 1.01$ and $0.09 < z < 0.11$. With 256 cells in both directions we have a resolution of 1280 cells per scale height. True resolution studies for convergence should eventually also consider micro physics, like molecular viscosity and thermal conductivity or realistic radiative processes appropriate in optically thick and thin regimes. Without these processes the range of unstable wavenumbers is unlimited and with higher resolution more unstable modes are possible.

\begin{figure*}
\gridline{\fig{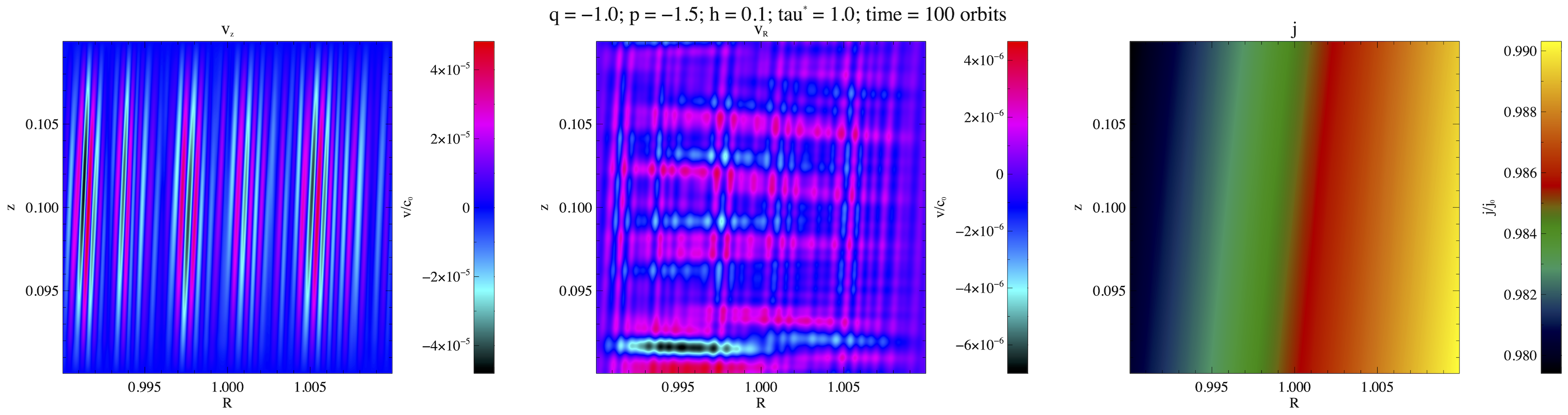}{0.9\textwidth}{(a): full model: $t = 101 \,\mathrm{ orbits}$}
}
\gridline{\fig{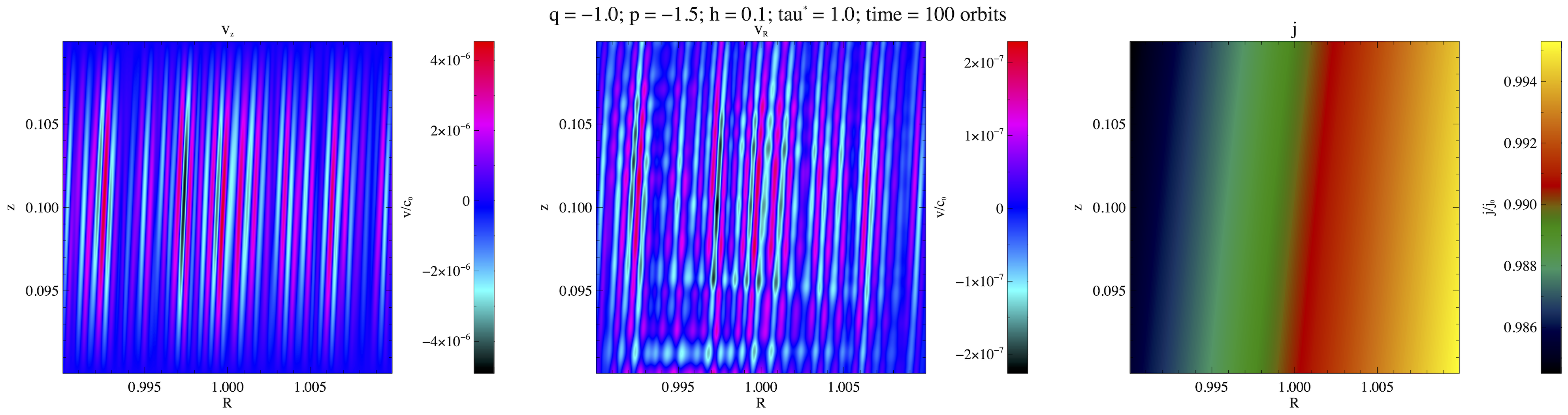}{0.9\textwidth}{(b): No COS $t = 100 \,\mathrm{ orbits}$}
}
\gridline{\fig{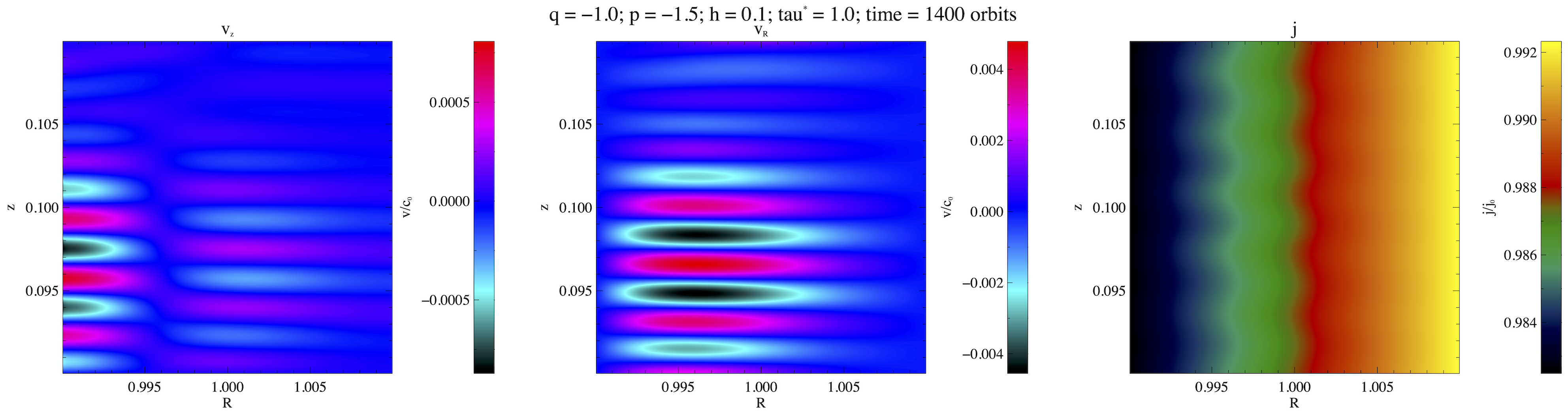}{0.9\textwidth}{(c): No GSF $t = 1400 \,\mathrm{ orbits}$}
}
  \caption{Simulation snapshots at $t = 100$ orbits for $q = -1, p = -1.5, \tau^* =1 $ at $z_0 = H$. In (a): the full model, (b): suppressed COS modes and (c): suppressed GSF modes. 
    \label{Fig:CV2ZQ1Z1T1_100}}
\end{figure*}

We conduct three different simulations for the same setup in terms of disk parameters $q,p,h,\tau^*$ but either use a "full" gravity model, with the correct conservative gravitational potential, a "No GSF" model, which suppresses vertical shear, and a "No COS" model, that has no radial temperature gradient, but vertical shear due to the modified gravity.
They all start from a $1\%$ perturbation in density, but no perturbation in pressure.

In Fig.\ \ref{Fig:CV2ZQ1Z1T1_100}, we compare the three simulations after 100 orbits. Both the full model and the "No COS" model show the prominent vertical GSF modes. The "No GSF" model shows the radial convective oscillations of COS. The absolute values of the velocities can be read from the time evolution of velocities in the three models in Fig.\ \ref{Fig:VMAXZQ1Z1T1}. We also produced a movie of the "No GSF" simulation (see: \href{https://youtu.be/xosb3Pll24A}{Movie2}), which displays the radial oscillations, which slowly move upward away from the midplane, while slowly growing in amplitude.

\begin{figure*}
\gridline{\fig{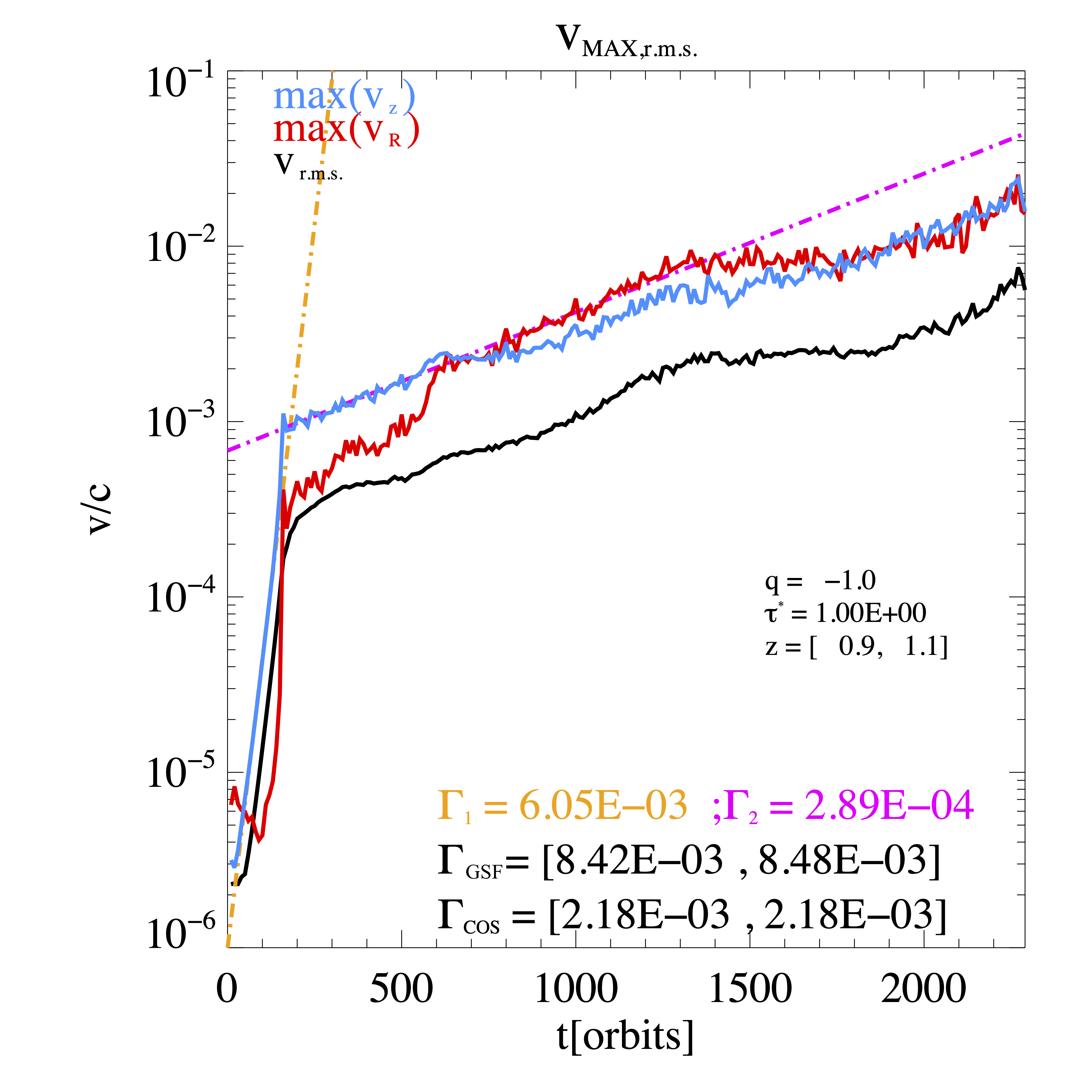}{0.3\textwidth}{(a): full model}
\fig{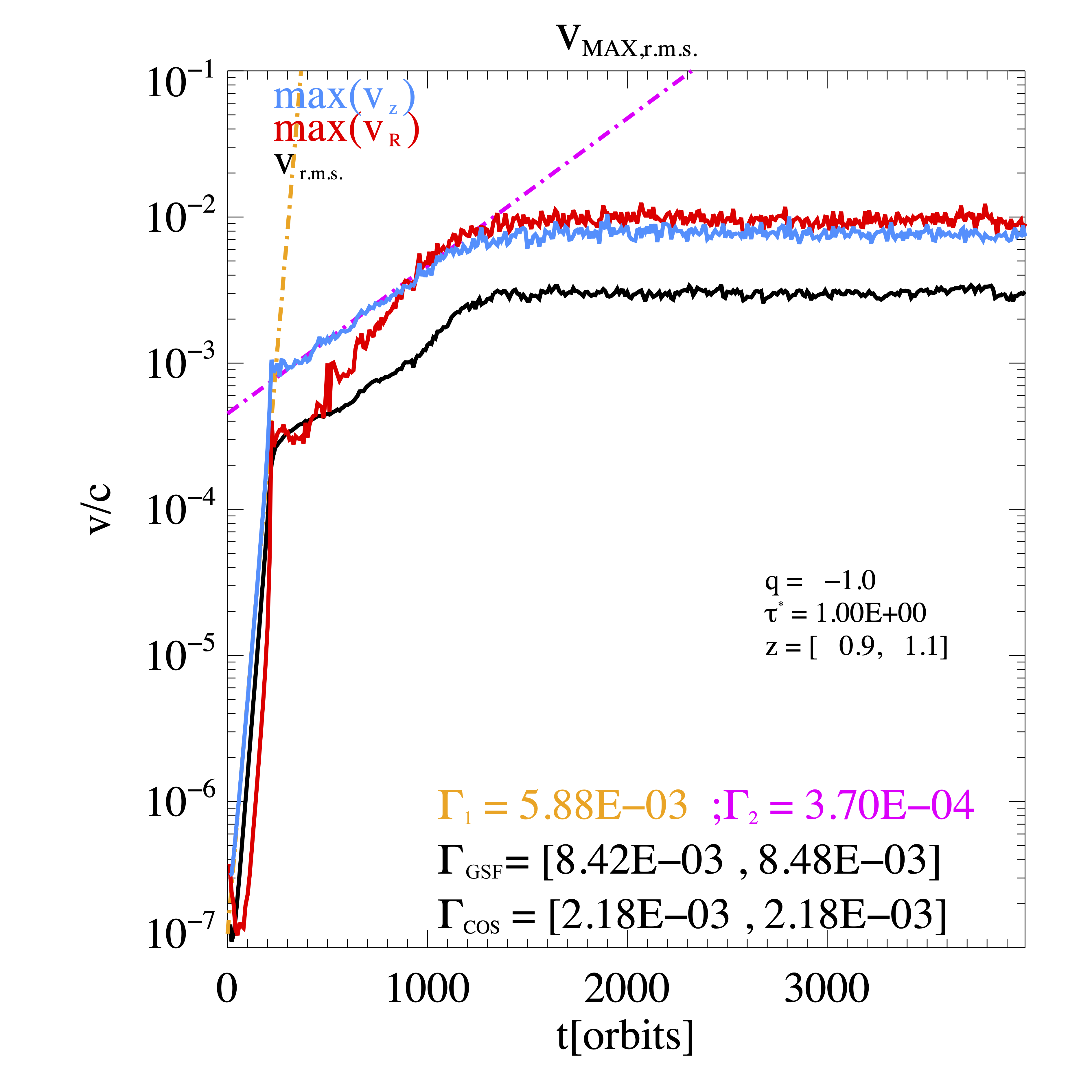}{0.3\textwidth}{(b): No COS}
\fig{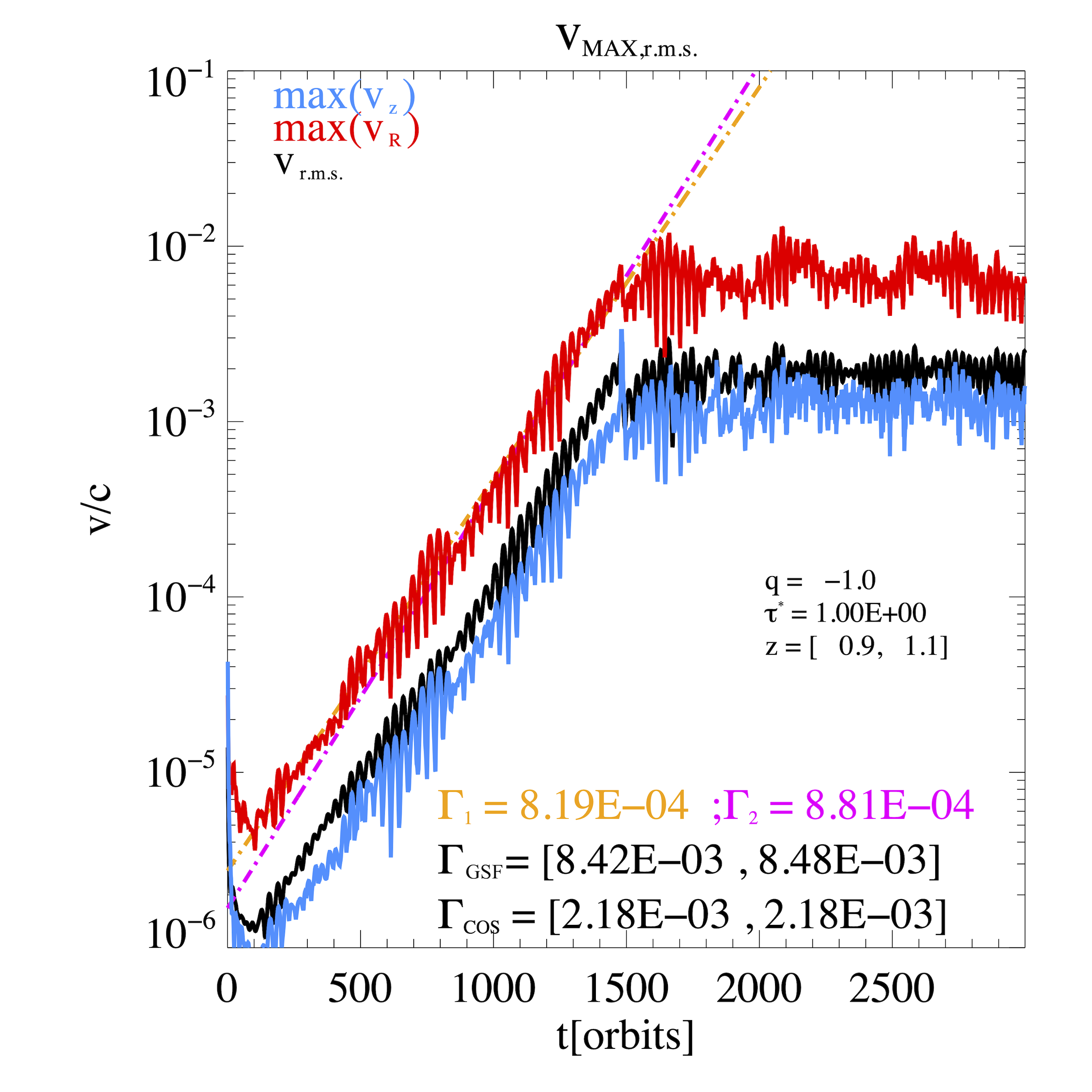}{0.3\textwidth}{(c): No GSF}
}
  \caption{Evolution of r.m.s.\ and maximal velocities for $q = -1, p = -1.5, \tau^* = 1$ for a small section of the atmosphere ($z = [0.9H, 1.1H]$) in units of the speed of sound and as function of time in units of orbits. In (a): full model (b): No COS: Convective modes are suppressed and (c): No GSF, suppressed GSF modes.
  We show the r.m.s. velocities (in the $R$-$z$ plane) as a solid black line. The largest radial velocity is indicated by the red line and the largest vertical velocity by a blue line. The yellow and magenta lines are the fitted growth rates $\Gamma_1$ and $\Gamma_2$ given in plot, along with the analytic growth rate range for the simulated domain.
    \label{Fig:VMAXZQ1Z1T1}}
\end{figure*}

Both models allowing for GSF (full and "No COS") reach the end of the linear growth phase after 200 orbits, at which time the r.m.s.\ velocity is about $3 \times 10^{-4} c_s$, whereas the pure COS simulation ("No GSF") grows for 1500 orbits to saturate at a level of $v_\mathrm{r.m.s.} = 2 \times 10^{-3} c_s$. At this amplitude, the COS channel modes create parasitic GSF modes (see Fig.\ \ref{Fig:CV2ZQ1Z1T1_1000}), same as shown in the spherical simulations close to the midplane in the previous section. 
{Note that the measured growthrates are about 2 times smaller than the analytic prediction, which may be related to the resolution, to the radial limitation of the computational domain, or possibly the artificial modification of the gravitational potential. In the full simulations in the previous section (See} Fig.\ \ref{Fig:VMAXQ1T1B}{), the measures COS growth rates matched the analytical derived estimation.}
The velocity amplitudes of the GSF runs continue to grow over time in a second growth phase, albeit at rates even longer than the expected COS modes. The "No COS" model stalls its growth at an amplitude of $v_\mathrm{r.m.s.} = 10^{-2} c_s$, but the full model continues to grow.

\begin{figure}
\centering
\gridline{\fig{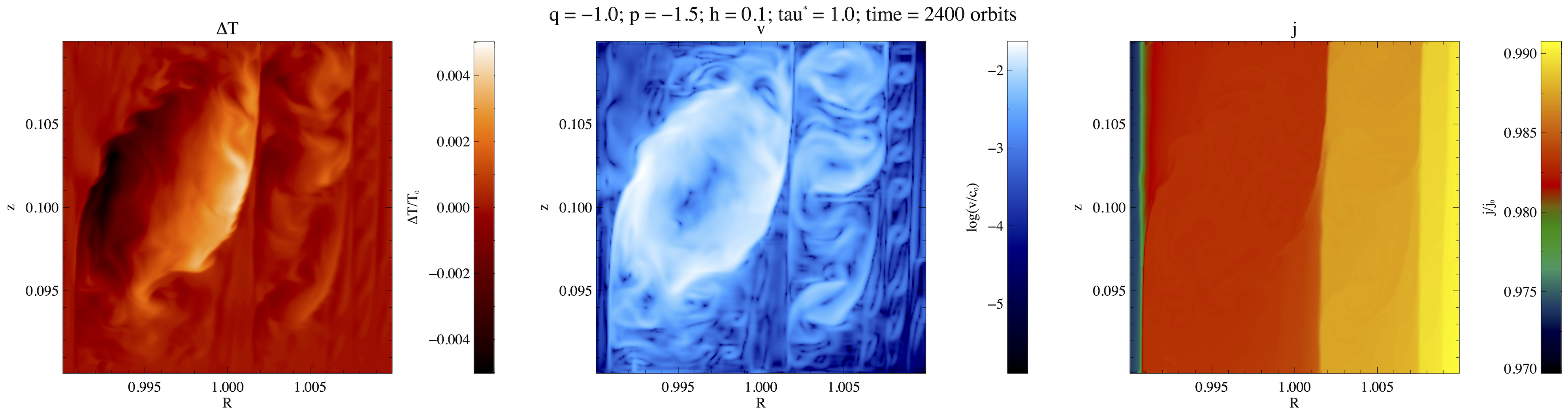}{0.9\textwidth}{(a): full model $t = 2400 \, \mathrm{ orbits}$}
}
\gridline{\fig{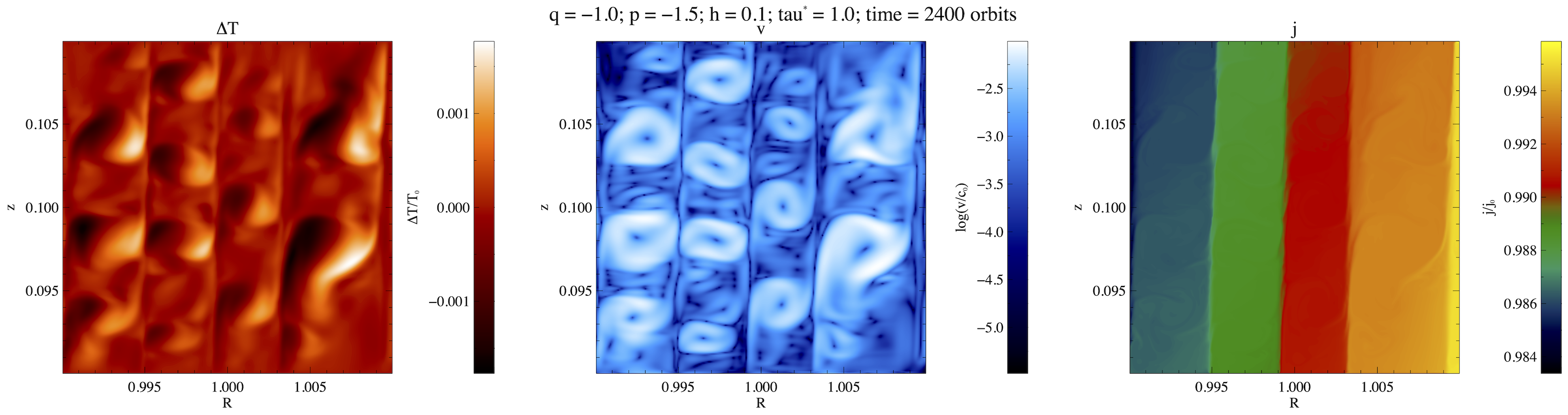}{0.9\textwidth}{(b): no COS $t = 2400 \,\mathrm{ orbits}$}
}
\gridline{\fig{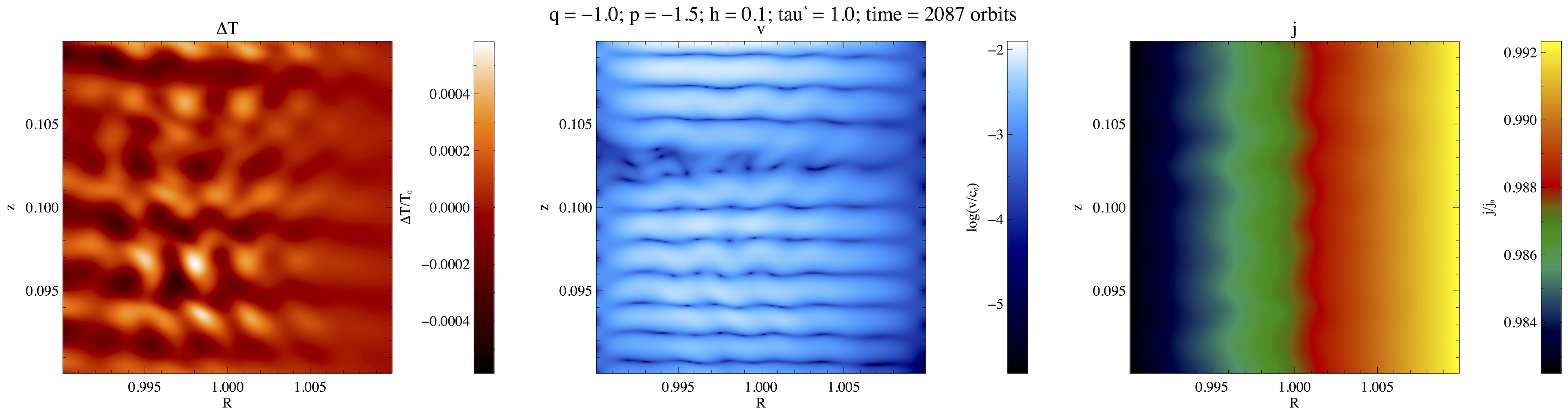}{0.9\textwidth}{(c): no GSF $t = 2087 \,\mathrm{ orbits}$}
}
  \caption{Simulation snapshots after the end of the individual linear growth phase for $q = -1, p = -1.5, \tau^* =1 $ at $z_0 = H$. Each row starting from the top: {(a) the full model, (b) suppressed COS modes, and (c) suppressed GSF modes}.
    \label{Fig:CV2ZQ1Z1T1_1000}}
\end{figure}

\subsection{Diagonal COS modes}
Our simulations with $q=-0.5$ did not show a clear linear growth of any modes, even though there should have been some unstable diagonal / slanted modes, albeit growing over rather long time scales. So do they exist and are only damped by the numerical scheme?

To test the hypothesis that in radial, stably stratified regions one can still have diagonal convective modes we chose a radial temperature profile of $q=-4$, which according to our equations should provide fast enough growth for COS modes. For a normal disk structure, this would be unstable to radial convection, so we choose an ad hoc density gradient of $p=4.4$. Thus as density increases radially, {like at the transition between an inner cavity and disk}, the radial pressure gradient always points inward for all heights considered, whereas the entropy will always decrease outward. It is not necessary to discuss if and where such a stratification would occur in a disk, it is only important that we show that for this case the predictions from the linear analysis can be tested. From this test we can infer about the validity of the dispersion relation for inclined modes in general. 
\begin{figure*}
\centering
\gridline{\fig{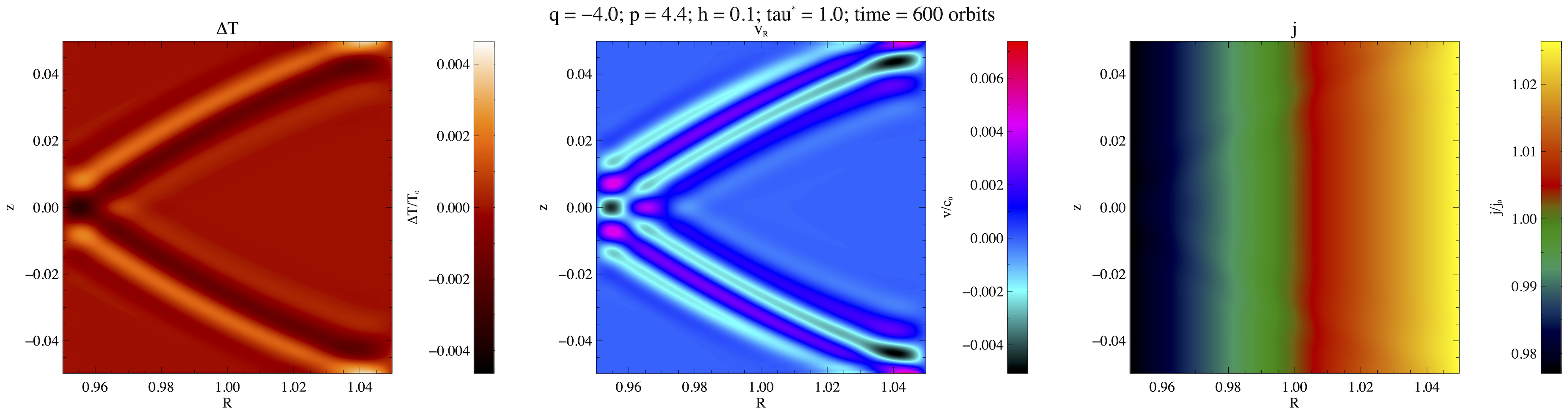}{0.9\textwidth}{No GSF $t = 600 \,\mathrm{ orbits}$}}
\gridline{\fig{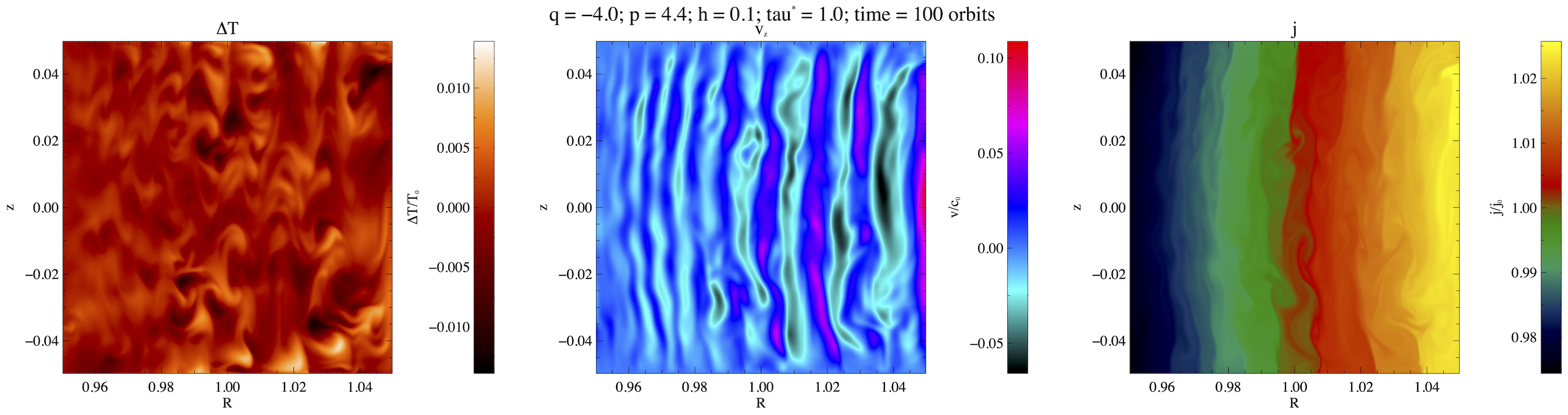}{0.9\textwidth}{Full Model $t = 100 \,\mathrm{ orbits}$}}
  \caption{Simulation snap shot during the linear growth phase for $q = -4, p = +4.4, \tau^* =1 $ at $z_0 = H$. When GSF is suppressed, one can clearly see the diagonal / slanted convection modes. In the full model GSF modes dominate.    \label{Fig:CV2ZQ4P4.4T1_300}}
\end{figure*}

We would have preferred to do such a test for the $q=-0.5$ models, but with our numerical scheme and resolution, we were not able to clearly identify the diagonal modes in these runs even for resolutions of 1024 cells per scale height. Possibly some other numerical hydrodynamic scheme will be able to do so eventually.

Our model uses the cylindrical setup, so we can test the full model and compare it with the model of suppressed GSF ("No GSF") 
using the modified gravity from the previous section. Our model has the dimensions in radial direction $0.95 < R < 1.05$ and vertical direction $-0.05 < z < 0.05$ using 512 cells each in both directions and the usual closed boundary condition. Disk aspect ratio $h=0.1$ and adiabatic index $\gamma = 1.4$ is the same as in all the other models, and the cooling time is $\tau^* = 1$. We perturb all three velocity components with random values in the range of $\pm 10^{-5}$ of the local speed of sound to allow for a fair comparison of the No GSF and full modes, which would respond differently to a perturbation in density.

In Fig.\ \ref{Fig:CV2ZQ4P4.4T1_300} we show a snapshot at $t=300$ during the linear growth phase. The COS modes appear as diagonal perturbations, which actually propagate orthogonal to their wave vector. The details of the evolution of these modes can best seen in a movie of this simulations (see  \href{https://youtu.be/mYtwbSXYACY}{Movie 3}). The growth rates are on the expected order of magnitude. The modes saturate after about 600 orbits at which time the flow becomes chaotic.

To show these diagonal modes was a matter of principle. If we do not suppress vertical shear, then as for all other parameters, the GSF will grow faster and eventually create bands of constant angular momentum with embedded eddies, at least for the axisymmetric setups considered here (see Fig.\ \ref{Fig:CV2ZQ4P4.4T1_300}). We will discuss the cause for the eddies some more in the following section.

\section{Horizontal Convection}
In basically all runs outside the midplane we found the formation of in-plane eddies. For short as well as long cooling times, as well as in the simulations of the previous section (see Fig.\ \ref{Fig:CV2ZQ1Z1T1_1000}), as long as the disk was baroclinic. Only the "No VSI" run showed no such eddies forming.

The reason for these in-plane whirls is two-fold. First, the vertical modes of GSF eventually create regions of constant angular momentum and then the baroclinic term is longer balanced by vertical shear (thermal wind), but produces vorticity.
In the fluid dynamics literature this effect is called horizontal convection (HC), which is not driven by a horizontal super adiabatic stratification, but directly by the baroclinic term. It occurs in systems with negligible rotation, which are baroclinic because of differential heating, for instance the sea-breeze effect \citep{Holton2012-kg}.

In the linear stage of our disk simulations $\kappa_z^2$ is balanced by the baroclinic term, but once the {vertical shear vanishes} we have $\kappa_z^2=0$. For the "thermal wind equation" we explicitly set radial and vertical velocities to constant zero $v_R = v_z = 0$, but now we can take the curl of the momentum equations and introducing the $\phi$ component of vorticity $\omega_\phi = \partial_R v_z - \partial_z v_R$ we find:
\begin{equation}
    \partial_t \omega_\phi = \frac{\left(\mathbf{\nabla}\rho \times \mathbf{\nabla}P\right)_\phi}{\rho^2}  + \kappa_z^2. 
\end{equation}
Thus in the initial state with vertical shear there is no horizontal convection possible. But as vertical shear and radial angular momentum gradient is locally removed, vorticity can be created, as the global rotation of the disk can be ignored. In that case we have a typical case of horizontal convection, the driving mechanism behind the Sea-breeze in geophysics \citep{Holton2012-kg}.

In global simulations one can observe the same effect, but additionally one finds that the shear between the vertical modes undergo a Kelvin-Helmholtz instability (see Melon Fuksmann A\&A, submitted). But these eddies rotate counter clock wise following the radial shear of vertical motions between the bands of constant angular momentum. 
Yet inside the bands of constant angular momentum the {eddies we observe in our simulations} are all rotating clockwise, as forced by the sign of the baroclinic term.
It seems that our local simulations saturate, before the KHI can be triggered. 

Like the COS and GSF, HC feeds on the baroclinicity of the disk, yet cannot be described by plane waves. For the COS, we consider oscillations that are amplified if the cooling time and the oscillation period are similar.
For the horizontal convection in a disk, there already have to be closed loops, i.e. closed streamlines of constant angular momentum, so the flow in the whirl can move freely. If we now perform an integral of $P \mathrm{d}V $ along this stream line, we can determine the work set free in the whirl, which in terms of density and pressure along the path $s$ is given by the integral
\begin{equation}
    W = \oint\limits_{s} P \frac{\partial V }{\partial s}\,\mathrm{d} s = - \oint\limits_{s} \frac{P}{\rho^2} \frac{\partial \rho}{\partial s}\,\mathrm{d} s.
\end{equation}
If we adopt the initial density and temperature structure for our disks, we can easily evaluate this integral numerically for arbitrary loop shapes and sizes. Using Stokes theorem, we can replace the line integral by the surface integral, where a clockwise circulation produces a positive area (and an anti-clock wise circulation a negative area)
\begin{equation}
    W = \int\limits_{S} \frac{\mathbf{\nabla}\rho \times \mathbf{\nabla}P}{\rho^2} \cdot d\mathbf{S} = -\int\limits_{S} \kappa_z^2 dS.
\end{equation}
Thus in the northern hemisphere with $\kappa_z^2 < 0$ for the initial equilibrium structure clockwise circulations are amplified, whereas in the southern hemisphere counter-clockwise whirls are amplified. 

For a barotropic system, {the integral of $P \mathrm{d}V $ for a closed loop} is always zero and thus no energy can be released, even though there may be changes in density and pressure along the path. But in a disk with a prescribed density structure of $p = -1.5$ and independent temperature structure $q =-1$, eddies will release energy when they rotate clockwise in the atmosphere above the midplane or counterclockwise in the atmosphere below the midplane. The {energy released by the eddy scales with its size} and with the velocity of the flow. If there was no thermal relaxation, then entropy would be conserved along the fluid line and the work integral would also be always zero. But for thermal relaxation towards said initial configuration, the atmosphere has a chance to maintain its baroclinicity. This is now already a difference with respect to the COS. The COS is bound by the epicyclic frequency and thus will operate poorly for larger cooling rates. But horizontal convection has no such restriction. The shorter the cooling time, the faster the gas can flow. Otherwise the cooling time will limit the velocity of the expanding gas.

The eddies on the other hand, sit in a baroclinic structure and thus they can be amplified by arbitrary short or long cooling times. Of course for long cooling times the velocity of the eddies will be limited by the cooling rate.

Checking Fig.\ \ref{Fig:CV2ZQ1Z1T1_1000}, we can clearly identify the eddies for the full model. Evaluating the integral $W$ along a stream line inside a whirl indeed results in a net release of thermal energy. But why does the pure COS model `No GSF' not show these modes? The integral of $p \mathrm{d} V$ would still suggest the release of energy, however in modifying the implementation of gravity we also have to integrate the release of potential energy in our non-conservative setup. And as gravity was modified to remove the vertical shear, the integral of gravitational forces exactly cancels the release of thermal energy and the baroclinic eddies vanish. The `No COS' simulation on the other hand, has a barotropic background and thus the integral of $p \mathrm{d} V$ vanishes. But we still see the same whirls at least initially as in the full simulation. In this case, the modified gravity used to generate vertical shear allows for the continuous release of potential energy as alluded to in the aforementioned Escher staircase analogy. We see that while it may still be valid to study the linear phase of GSF and COS in the modified gravity regime, the non-linear phase will contain some non-physical results.

So at least in the full version of our axisymmetric simulations, we can clearly identify these baroclinic driven eddies, which are responsible to increase the r.m.s.\ velocity after the saturation of the GSF modes by an order of magnitude. All whirls rotate clockwise, as the counterclockwise whirls would convert motion into heat like a heat engine. In the southern hemisphere, below the midplane, of course the counterclockwise eddies will be the ones amplified.

As mentioned before, these whirls are only so efficient because they sit in a band of constant angular momentum, which itself would be unstable in 3D simulations. Thus the possibility of having some baroclinic driven eddies will have to be studied in dedicated 3D simulations in the future. But it is clear that any developing non-linear flow in 3D has the chance to tap into the baroclinicity of the disk, they don't have to be axisymmetric to release energy. One will have to hunt for these modes in future simulations.

A similarity also exists between the baroclinic driven eddies and the vortex amplification in the subcritical baroclinic instability (SBI) \citep{Petersen2007a,Petersen2007b,Lesur2010}, in the sense that for both instabilities it needs a pre-existing vortex. For SBI, this is a vortex in the $R$-$\phi$ plane of the disk and for the "horizontal convection" it is in the $R$-$z$ plane of the disk. But whereas the "horizontal convection" operates on a background state that is already baroclinic and thus an amplification will happen for arbitrarily small velocities, an SBI vortex needs a certain rotation velocity to generate a $P(\rho)$ structure that can deliver work $W$ because of the delay of cooling and heating during the rotation, i.e.\ gas moving outward is warmer and thus lower in density than the inward moving gas. For vanishing cooling times, $P$ and $\rho$ is symmetric with respect to inward or outward motion and no work can be released $W=0$. The optimal cooling time once again is given by the rotation frequency of the SBI vortices \citep{Raettig2013} very much like for the COS modes.

Baroclinic driven { vortices (or eddies)} on the other hand will thrive at the fastest cooling rates, because then the work integral will remain positive even for the largest velocities. The faster the velocities of the eddies (in two or or three dimensions) will be, the more energy can be released {per unit time}. In that sense, the eddies may be a robust feature in disks once it is triggered, yet it remains to be seen if it occurs and plays a role in three-dimensional simulations of disks.

\section{Conclusion}
In Paper I, we studied the linear stability of vertically isothermal gas disks with a radial temperature gradient $q$ for finite {thermal relaxation timescales $\tau$}.
For $q \ne 0$ and $\tau < \infty$, we always found linearly unstable solutions in our stability analysis. In the present paper, the predicted growth rates could be confirmed in nonlinear simulations with reasonable agreement, considering the dissipation inherent in our numerical scheme. 

Protoplanetary disks around young stars possess some temperature gradient and a finite thermal relaxation, so they can adapt their temperature to irradiation and possible internal processes. In light of our findings, this implies that there cannot be a stable rotation profile or likewise stable density structure for these disks. The instabilities arise from the baroclinic structure of the disks, which is responsible for vertical shear (aka thermal wind) and convectively unstable stratification in selected directions.

The effect of baroclinic stratifications to generate vertical shear was already considered for GSF and VSI instabilities, albeit in the regime of sufficiently small cooling times. We tested now the predictions for the growth of GSF modes for arbitrary cooling times. The stability criterion of VSI and GSF are identical, yet GSF growth rates for long cooling times scale proportionally to the square of the temperature gradient $q^2$ and inversely with cooling time, whereas VSI and GSF for short cooling times scales linearly with $q$, which we confirm in our numerical simulations.

In the long cooling time regime, there are always COS modes with growth rates at least smaller by a factor of two with respect to GSF. In the midplane we observe that both modes grow simultaneously, with the GSF having a faster start but also earlier saturation. In the long run, the COS takes over and eventually dominates the dynamics.

In the atmosphere, the COS modes are hardly visible in our simulations, and only by suppressing the GSF could we study the evolution of COS modes, confirming their predicted growth rates for $z=H$.

For normal $q=-0.5$ we found GSF modes in the atmosphere, but the identification of COS modes was unsuccessful, for which we blame the numerical dissipation. Yet, for an artificially boosted temperature gradient $q=-4$ counterbalanced by an inverse density gradient $p=+4.4$ we could show that, even if the radial structure is convectively stable, there will be the predicted diagonal / slanted convection modes.

We can thus confirm that the radial density gradient in the midplane of the disk has neither a direct influence on the growth rates of GSF nor on those of COS. Specifically, the sign of the radial entropy gradient {does not affect the onset of} COS.

Besides the confirmation of the growth rates of unstable modes as derived in Paper I we also identified a new mode of instability.
{Saturation in our local axisymmetric modes occurs when bands of constant angular momentum form. In these bands neither GSF nor COS can operate, as there is no vertical shear and the epicyclic frequency vanishes. Also the stable direction of buoyancy $N^2_+ \approx N^2_z$ dominates over the unstable direction $N^2_+ + N^2_- > 0$ and thus convection would be quenched into thin sheets and not explain the circular eddies we find.
Still we observe the formation and amplification of eddies in these bands that are all rotating in the direction defined by the baroclinicity.
As also discussed in (Melon Fuksman et al. A\&A, submitted) it is the baroclinic term that drives the eddies. The process is therefore described best as horizontal convection or "Sea-breeze" effect \citep{Holton2012-kg}. 
}
The stronger the baroclinic term is enforced via thermal relaxation, the more energy can be pumped into the system. But these preliminary results are possibly restricted to our axisymmetric setup and thus it is an open question whether they also exist in full three-dimensional simulations, or whether at least similar nonlinear flow features in three-dimensional simulations can also tap into the baroclinic energy reservoir.

For the cases studied in this paper, i.e. vertically isothermal, Newtonian cooling with a fixed cooling time for all wave numbers, no viscosity, and strict axisymmetry, we find that GSF always grows faster than the COS modes, but at least sometimes it saturates earlier. {A situation in which additional effects as for instance the sedimentation of dust leads to an additional stabilization of the vertical stratification \citep{Lin2019} could create a situation in which COS may dominate over GSF modes, as already discussed in \citet{Shibahashi1980} and \citet{Tassoul2000}.}

Three-dimensional simulations will be necessary to see if we can also reproduce the linear growth rates if axisymmetry is not artificially constrained. We can expect that saturation will not occur when bands of constant angular momentum form as they themselves are unstable to non-axisymmetric modes (Rayleigh criterion).

Then we will see if a similar baroclinic driving of turbulence as we see it in the present simulations does also occur in the nonlinear state of fully three-dimensional turbulence similar to the SBI as reported in \citep{Petersen2007a,Petersen2007b,Lyra2011,Raettig2012}, which were all either vertically integrated or vertically unstratified models and thus needed cooling times on the order of the orbital period.

It will be interesting to measure how much the emerging three-dimensional turbulence can draw energy from the baroclinic state of the disk even in the short cooling time regime.

With sufficient resolution, it will be possible to perform simulations in the regime of short cooling times $\tau < \tau_c$ where VSI and GSF will dominate, but also in the long cooling time regime $\tau > \tau_c$ with COS and GSF of similar strength. A third regime for very long cooling times ($\tau > 10 \Omega^{-1}$) may then show the transition from thermal instabilities to ZVI \citep{Barranco2018}.

Based on our numerical experiments we confirm that all disks around young stars are hydrodynamically unstable. Protoplanetary disks all have a temperature gradient for most of their radius, generated by stellar irradiation and thus are all baroclinic and prone to GSF and COS. But the growth rates of GSF and COS and their interaction with other instabilities have to be considered now. Ideal MRI would easily outgrow the hydrodynamic instabilities by far, as we showed for the case of SBI \citep{Lyra2011} and studies investigating to what extent non-ideal MHD regime will allow for hydrodynamic instabilities of the VSI nature have just been started recently \citep{Latter2022}. A full picture of MHD and pure HD effects in disks is an ambitious goal, yet now we have the tools and know the necessary resolution, at least from the pure hydrodynamic perspective.

\section*{Acknowledgments}
The authors wish to thank Natascha Manger, Orkan Umurhan, and Wladimir Lyra for providing feedback on early versions of this manuscript. H.K. and J.D.M.F. are supported by the German Science Foundation (DFG) under the
priority program SPP 1992: “Exoplanet Diversity” under
contract KL 1469/16-1/2. HB acknowledges the support of the NASA Theoretical and Computational Astrophysics Networks (TCAN) award 80NSSC19K0639. Simulations were performed on the ISAAC and VERA clusters of the MPIA and the COBRA, HYDRA and RAVEN clusters of the Max-Planck-Society, both hosted at the Max-Planck Computing and Data Facility in Garching (Germany). H.K. also acknowledges additional support from the DFG via the Heidelberg Cluster of Excellence STRUCTURES in the framework of Germany's Excellence Strategy (grant EXC-2181/1 - 390900948).

\appendix
\section{Movies}
We produced a set of movies from our simulations, as we find it very insightful to "see" how the instabilities develop. All movies show four panels, yet the shown information can vary, depending on what we have to highlight.
\begin{figure}
    \centering
    \includegraphics[width=0.8\linewidth]{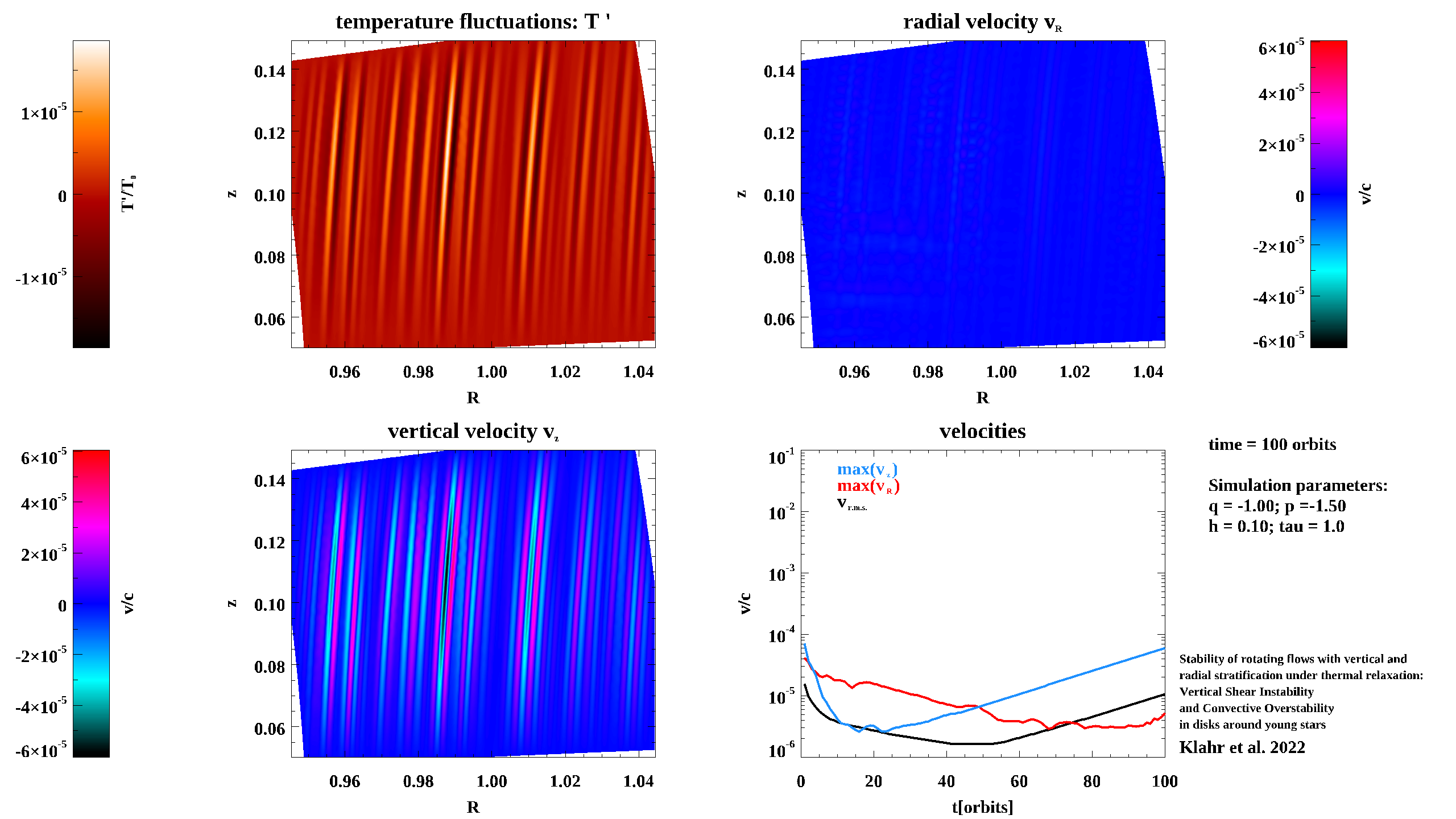}
    \caption{Snapshot from Movie 1a at $t = 100$. Here we show radial and vertical velocity, together with the temperature deviation from the background and the evolution of r.m.s. velocity as well as velocity maxima.}
    \label{fig:100}
\end{figure} 
\begin{figure}
    \centering
    \includegraphics[width=0.8\linewidth]{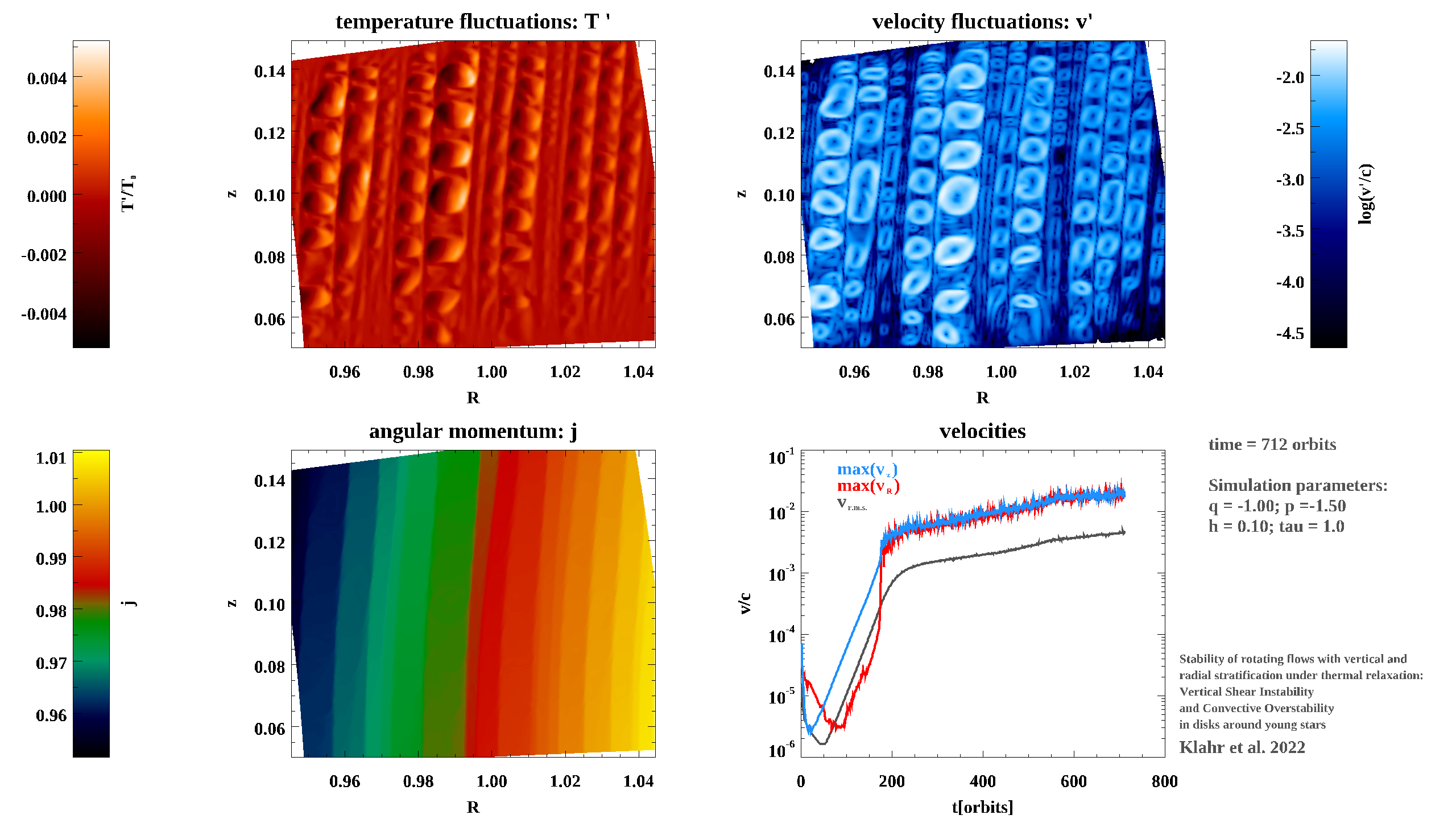}
    \caption{Snapshot from Movie 1a at $t = 712$. Here we show the absolute velocity to highlight the formation of the eddies, together with the specific angular momentum, the temperature deviation from the background and the evolution of r.m.s. velocity as well as velocity maximal}
    \label{fig:712}
\end{figure} 

The first panel shows the local relative deviation from the background temperature $T' = \frac{T-T_0}{T_0}$ (see Figure \ref{fig:100} and \ref{fig:712}). 
The range of the color scheme expands as the amplitude increases with time. 
The second plot represents either the radial velocity (see Figure \ref{fig:100}) or the mass flux in the $R,\theta$ plane (see Figure \ref{fig:712}) ,  i.e.\ $\rho v' \propto \rho \sqrt{v_R^2 + v_z^2}$, scaled on the speed of sound $c=h=0.1$ and density $\rho_0 = 1$ in the midplane at $R = 1$, to emphasize the velocities in regions of higher density, especially the case where there are many scale heights in the vertical direction. The lighter the blue, the higher the flow velocities. Thus the white streaks follow the {streamlines} in the simulation, indicating the dominant directions for the gas flow velocities. 
The third plot shows either the vertical velocity (see Figure \ref{fig:100}) or the distribution of specific angular momentum $j = \Omega R^2$ (see Figure \ref{fig:712}). 

The fourth plot shows the evolution of the r.m.s.\ velocity (black) as well as the largest radial (red) and vertical velocities (blue), in a region around the center of the simulation, which is radially and vertically half as wide as the respective simulation.

In total we present 4 Movies.
Movies 1a and 1b are for the $q=-1$ and $\tau^* = 1$ cases, 2a for the upper layer and 2b for the midplane as discussed in section 2.1. In Movie 1a, which correspond to Figures 7, 8 and 9, one can observe how initially the typical vertical GSF modes are growing and after the saturation the formation of the eddies of constant angular momentum, which are the sign for the non linear symmetric instabilty, tapping directly in the baroclinic state of the disk via the $PdV$ term. Movie 1b covers a simulation placed in the midplane. Here first GSF starts growing, then later COS takes over. This is the same simulation shown in Figures 10 and 11

Movie 2 is a cylindrical version of the simulation shown in Movie 1a, but now in cylindrical coordinates and suppressing the GSF as discussed in section 2.2, thus the same simulation as depicted in Figure 13. One can observe the amplification and vertical drift of the COS modes that also go nonlinear, once the vertical shear in the radial oscillation alows for a parasitic version of the GSF. 

Movie 3, see section 2.3 for details, finally suppresses the GSF and is located around a radial stable stratified midplane with $q=-4$ and $p=4.4$. This movie shows the development of the diagonal convective modes, aka slanted convection.
\begin{deluxetable}{lllll}[tb!]
\tabletypesize{\footnotesize}
\tablecaption{Movies:\label{tab:Movies} All models use $h = 0.1$ and $\tau^* = 1$}
\tablehead{\colhead{Name} &\colhead{coordinates} & \colhead{dimensions} &\colhead{parameters} & \colhead{Description}} 
\startdata
\href{https://youtu.be/YeB_fOlLwsE}{Movie 1a} & spherical &$ 0.5 H < z < 1.5 H$ &  $q= -1; p = -1.5$ & GSF above midplane\\
\href{https://youtu.be/02TnKk5rNv0}{Movie 1b} & spherical & $ -0.5 H < z < 0.5 H$ &  $q= -1; p = -1.5$ & GSF and COS in midplane\\
\href{https://youtu.be/r8_XBtY9uE4}{Movie 2} & cylindrical &$ 0.5 H < z < 1.5 H$ &  $q= -1; p = -1.5$ & only COS modes above midplane\\
\href{https://youtu.be/mYtwbSXYACY}{Movie 3} & cylindrical &$ -0.5 H < z < 0.5 H$ &  $q= -4; p = +4.4$ & only COS in midplane\\
\enddata
\end{deluxetable}


\bibliographystyle{aasjournal}

\end{document}